\documentclass[aps,prl,twocolumn,superscriptaddress,10pt]{revtex4-1}
\usepackage[utf8]{inputenc}
\usepackage{amsmath}
\usepackage{amssymb}
\usepackage[normalem]{ulem}
\usepackage{graphicx}
\usepackage{pdfpages}
\makeatletter
\AtBeginDocument{\let\LS@rot\@undefined}
\makeatother

\usepackage{color}
\usepackage{verbatim}
\usepackage{booktabs}
\usepackage{float} 
\usepackage{listings}
\usepackage{alltt}

\usepackage{hyperref}
\usepackage{cleveref}
\usepackage{subfigure}

\usepackage{silence}
\usepackage[normalem]{ulem}

\crefname{equation}{Eq.}{Eqs.}
\Crefname{equation}{Equation}{Equations}
\crefname{figure}{Fig.}{Figs.}
\Crefname{figure}{Figure}{Figures}
\crefname{section}{Sect.}{Sects.}
\Crefname{section}{Section}{Sections}

\newcommand{\ket}[1]{| #1 \rangle}

\newcommand{\expo}[1]{\text{e}^{ #1 }}

\newcommand{\sx}{\hat{\sigma}_x}

\newcommand{\ha}{\hat{a}}
\newcommand{\had}{\hat{a}^\dagger}
\newcommand{\hf}{\hat{f}}
\newcommand{\hfd}{\hat{f}^\dagger}

\newcommand{\hH}{\hat{H}}
\newcommand{\wc}{\omega_r}

\newcommand{\szo}{\hat{\sigma}_{z1}}
\newcommand{\szt}{\hat{\sigma}_{z2}}
\newcommand{\szi}{\hat{\sigma}_{zi}}

\newcommand{\mEp}{\mathcal E_p}

\usepackage{ulem}

\begin{document}

\title{Qubit Parity Measurement by Parametric Driving in Circuit QED}
\author{Baptiste Royer}
\affiliation{Institut quantique and D\'epartment de Physique, Universit\'e de Sherbrooke, 2500 boulevard de l'Universit\'e, Sherbrooke, Qu\'ebec J1K 2R1, Canada}
\author{Shruti Puri}
\affiliation{Department of Applied Physics, Yale University, PO BOX 208284, New Haven, CT 06511}
\author{Alexandre Blais}
\affiliation{Institut quantique and D\'epartment de Physique, Universit\'e de Sherbrooke, 2500 boulevard de l'Universit\'e, Sherbrooke, Qu\'ebec J1K 2R1, Canada}
\affiliation{Canadian Institute for Advanced Research, Toronto, Canada}


 \begin{abstract}
Multi-qubit parity measurements are essential to quantum error correction. Current realizations of these measurements often rely on  ancilla qubits, a method that is sensitive to faulty two-qubit gates and which requires significant experimental overhead. We propose a hardware-efficient multi-qubit parity measurement exploiting the bifurcation dynamics of a parametrically driven nonlinear oscillator. This approach takes advantage of the resonator's parametric oscillation threshold  which is a function of the joint parity of dispersively coupled qubits, leading to high-amplitude oscillations for one parity subspace and no oscillation for the other. We present analytical and numerical results for two- and four-qubit parity measurements with high-fidelity readout preserving the parity eigenpaces. Moreover, we discuss a possible realization which can be readily implemented with the current circuit QED experimental toolbox. These results could lead to significant simplifications in the experimental implementation of quantum error correction, and notably of the surface code.
 \end{abstract}

\maketitle


\section{Introduction}
Quantum error correction (QEC) protects fragile quantum information from decoherence and will play a vital role in large-scale quantum computations. 
Typical QEC codewords are defined in a given eigenspace of multiple parity operators. When an error occurs, the state of the qubits leaves the codespace, something that is revealed by measuring the parity operators.
Since these measurements have to be performed repeatedly, it is crucial that they be of high fidelity. Moreover, to avoid introducing errors, these measurements should leave the parity subspaces intact, \textit{i.e.} states within a given parity subspace should remain unperturbed by the measurement. 

In practice, parity measurement strategies can be broadly classified as direct or indirect. The latter approach, used in recent demonstrations of small-scale error correction~\cite{Barends:14a,Corcoles:15a,Kelly:15a}, relies on a series of two-qubit entangling gates between the data qubits and an additional ancilla qubit which is subsequently measured~\cite{Pfaff:13a,Chow:14a,Saira:14a,Barends:14a,Corcoles:15a,Kelly:15a,Takita:16a}.
Drawbacks of this strategy are the accumulation of errors due to faulty two-qubit gates and the experimental overhead which could become an impediment to the implementation of larger QEC codes.

Faulty gates and overhead issues can be addressed by using \emph{direct} parity measurements. The central idea in this approach is to map the parity information onto the state of a common mode coupled to the data qubits and which is then measured. For example, a possible strategy to realize direct measurements of two-qubit parity in circuit quantum electrodynamics (QED) is by monitoring the response of a resonator dispersively coupled to the qubits. In this situation, the frequency of the oscillator, and therefore its response to a drive, becomes dependent on the joint-qubit parity~\cite{Hutchison:09a,Lalumiere:10a,Riste:13a}.
A challenge with this method is to design and implement a protocol which preserves the parity eigenspaces. In other words, in an ideal parity measurement, the common mode and its environment gain information \emph{only} about which parity subspace (even or odd) the qubits state belongs to. 
Possible improvements to overcome this eigenspace dephasing were introduced in Refs.~\cite{Tornberg:10a,Frisk-Kockum:12a,Govia:15a,Huembeli:17a}, but require quantum-limited amplifiers with unit efficiency~\cite{Tornberg:10a,Frisk-Kockum:12a} or high-efficiency single microwave photon detectors~\cite{Govia:15a,Huembeli:17a}. 

Here, we introduce a scheme for direct, high-fidelity parity measurements that leaves the parity subspaces intact. Our approach is based on dispersively coupling multiple qubits to a nonlinear resonator driven by a two-photon parametric pump. This situation leads to a qubit parity-dependent parametric oscillation threshold. 
When the qubits are in the even subspace, the amplitude of the two-photon drive is below the parametric oscillation threshold and the resonator state remains close to vacuum. On the other hand, in the odd subspace, the parametric drive is above threshold and the resonator bifurcates to a high-amplitude state.
We show that by monitoring the amplitude of the resonator output field with standard homodyne detection, it is possible to infer the parity of the qubit ensemble with high fidelity while preserving both even and odd parity subspaces. 
Importantly, we show that the photon number in the high amplitude state can be increased by reducing the resonator nonlinearity, leading to an increased signal-to-noise ratio (SNR) at constant eigenspace dephasing.
These ideas are generalized to more than two qubits by using a multi-tone parametric drive targeting the multiple dispersive shifts corresponding to the same parity subspace.

These ideas can be applied to different types of qubits coupled to oscillators. For concreteness, here we present a circuit QED implementation~\cite{Blais:04a,Wallraff:04a} based on transmon qubits~\cite{Koch:07a} that can be easily implemented with the current circuit QED toolbox~\cite{Krantz:16a,Boutin:17b}.

\section{Results and Discussion}


\subsection{Parametrically Driven Non-linear Resonator}
Before introducing our proposal for multi-qubit parity measurements, we present its main component: a resonator of frequency $\wc$ and Kerr non-linearity $K$. In the presence of a resonant parametric two-photon drive 
$\mEp$ of frequency $\omega_p = 2\wc$
and in a frame rotating at $\wc$, this system is described by the Hamiltonian ($\hbar = 1$)
\begin{equation}\label{eq:ham}
\hH_R = \frac{\mEp}{2} (\ha\ha + \had\had) - \frac{K}{2} \had\had\ha\ha,
\end{equation}
where $\ha$ and $\had$ denote the resonator's annihilation and creation operators, respectively. When the drive is turned off, $\mEp = 0$, the steady state of the system is the vacuum state.
Below the parametric oscillation threshold, $\mEp < \kappa/2$ with $\kappa$ the single-photon loss rate of the resonator, this system corresponds to the widely used Josephson Parametric Amplifier (JPA)~\cite{Boutin:17b} with a vacuum-squeezed steady state.
Above $\mEp > \kappa/2$, this system bifurcates into one of two states of equal amplitude but opposite phases characterized by $\langle \ha \rangle_{\mathrm{ss}} = \pm \alpha_\mathrm{o}$ with~\cite{Wustmann:13a,Puri:17a}
\begin{align}\label{eq:alpha0amplitude}
|\alpha_\mathrm{o}| &= \left(\frac{\mEp^2 - \kappa^2/4}{K^2}\right)^{1/4},\\ \label{eq:alpha0phase}
\theta_\mathrm{o} \equiv \text{Arg}[\alpha_\mathrm{o}] &= \frac{1}{2} \tan^{-1}\left(\frac{\kappa}{\sqrt{4\mEp^2-\kappa^2}} \right).
\end{align}
 Since both the Hamiltonian $\hH_R$ and the dissipation are symmetric under the transformation $\ha \rightarrow -\ha$ (see Methods), in steady state the resonator occupies either of the two states with equal probability, leading to a null \emph{average} displacement of the resonator field. However, a single shot homodyne measurement of the resonator steady state will always reveal a large amplitude $|\alpha_\mathrm{o}|$. Once the resonator has latched onto one of its two steady states, tunnelling to the other is highly suppressed for large values of $|\alpha_\mathrm{o}|$~\cite{Wielinga:93a, Marthaler:06a, Puri:17a}. In the limit where the two-photon drive is well above the parametric oscillation threshold $\mEp\gg\kappa/2$, the two steady states are coherent states. 

If the parametric drive is detuned such that $\wc-\omega_\textrm{p}/2=\delta$, the system Hamiltonian becomes
\begin{equation}\label{eq:ham_nr}
\hH_{R,\delta} = \delta\had\ha+\hH_R.
\end{equation}
At large detunings $\delta^2>\mEp^2 - \kappa^2/4$, the vacuum-squeezed state is a steady state of the system, with the squeezing axis governed by the sign of the detuning $\delta$~\cite{Puri:17a}. The degree of squeezing decreases as the ratio $|\delta|/\mEp$ increases and, for $|\delta|\gg\mEp$, the steady state is very close to the vacuum state.

\begin{figure*}[!t]
\centering
\includegraphics[scale = 1]{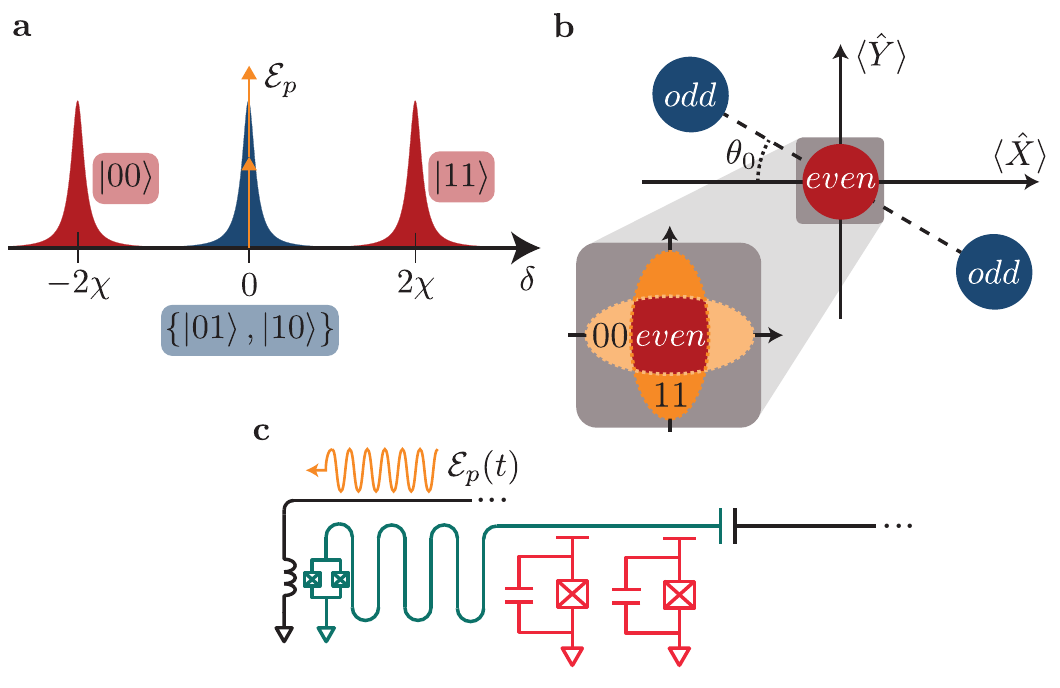}
\caption{\textbf{a} Qubit-state dependent frequency of the resonator. The parametric two-photon drive (orange) is resonant when the qubits are in the odd subspace, $\delta = 0$ (blue), and strongly detuned when the qubits are in the even subspace, $\delta = \pm 2\chi$ (red).
\textbf{b} Resonator phase space under two-photon driving. In the odd subspace, the resonator bifurcates in either $\pm \alpha_\mathrm{o}$ (blue), while in the even subspace it stays close to vacuum (red). The qubit parity is inferred by monitoring the amplitude of the field leaking out of the resonator. Inset: In the qubit even subspace, fluctuations are increased in a qubit-state dependent quadrature, leading to slow dephasing inside the subspace. \textbf{c} Possible circuit QED realization of the two-qubit parity measurement. Transmon qubits (red) are capacitively coupled to an off-resonant, non-linear resonator (green).}
\label{fig:schematic}
\end{figure*}


\subsection{Two-qubit Parity Measurement}
We now turn to the core of our proposal, first considering two-qubit parity measurements. More precisely, we aim to distinguish the odd subspace spanned by the two-qubit states $\{\ket{01},\ket{10}\}$ from the even subspace spanned by $\{\ket{00},\ket{11}\}$. To this end, we take two qubits dispersively coupled with equal strength $\chi$ to the parametrically driven nonlinear resonator. In a frame rotating at $\wc$, this system is described by the Hamiltonian
\begin{equation}\label{eq:ham2qb}
\hH_{2qb} = \chi (\szo + \szt)\had \ha + \hH_R,
\end{equation}
where $\szi$ is the Pauli Z operator for the $i^{th}$ qubit. Under this dispersive coupling, the resonator frequency becomes qubit-state dependent. 
We note that single-qubit readout in a similar setup was proposed in Ref.~\cite{Wustmann:13a} and experimentally demonstrated in Ref.~\cite{Krantz:16a}.

The above Hamiltonian, combined with the discussion of the previous section, immediately suggests an approach for multi-qubit parity measurement. Indeed, in \cref{eq:ham2qb}, the qubits induce a dispersive shift of the resonator frequency that will change the parametric oscillation threshold of the two-photon pump in a parity-dependent manner.
More precisely, if the state of the qubit lies in the odd subspace, $\ket{\psi_\mathrm{o}} = c_{01}\ket{01}+c_{10}\ket{10}$, the two dispersive shifts cancel as illustrated in \cref{fig:schematic}\textbf{a}. With $\delta = 0$, the system then behaves as a resonantly driven nonlinear resonator. Consequently, in the odd subspace, the resonator bifurcates to a large amplitude state as illustrated in \cref{fig:schematic}\textbf{b}. The combined qubits-resonator system thus evolves from the initial state, $\ket{\Psi(0)} = \ket{\psi_\mathrm{o}}\otimes\ket{0}$, to one of the two steady state $\ket{\Psi(t)} = \ket{\psi_\mathrm{o}}\otimes\ket{\pm \alpha_\mathrm{o}}$. Importantly, the phase of the oscillations, $\text{Arg}[\langle \ha \rangle_\textrm{o}] = \theta_\mathrm{o}, \theta_\mathrm{o} + \pi$, is independent of the state of the qubits within the odd subspace. In this situation, monitoring the output field of the resonator using standard homodyne measurement of the $X_{\theta_\mathrm{o}} = \left<\ha \expo{-i \theta_\mathrm{o}} + \had \expo{i \theta_\mathrm{o}}\right> $ quadrature reveals a large photon population in the resonator, $|\langle\ha\rangle_\textrm{o}|^2=|\alpha_\mathrm{o}|^2$. Note that during the homodyne measurement, the field can in principle switch between the two steady states $\pm \alpha_\mathrm{o}$, something that can reduce the measurement fidelity. However, these switching events are rare for large $|\alpha_\mathrm{o}|$~\cite{Wielinga:93a, Marthaler:06a, Puri:17a}.

On the contrary, in the even subspace, $\ket{\psi_\mathrm{e}} = c_{00}\ket{00}+c_{11}\ket{11}$, the dispersive shifts of the two qubits add up and the two-photon drive is off-resonant by $\delta = \pm 2\chi$. For dispersive shifts $|2\chi| \gg \sqrt{\mEp^2 - \kappa^2/4}$~\cite{Wustmann:13a}, the vacuum state remains a stable steady state even after activation of the two-photon drive as schematically depicted in \cref{fig:schematic}\textbf{b}. That is, the system remains in the initial state, $\ket{\psi_e}\otimes\ket{0}$. In this case, tracking the output of the resonator with homodyne measurement results in a null amplitude $|\langle\ha\rangle_\textrm{e}|=0$.

In practice, because the dispersive shifts are finite, the resonator state will deviate from vacuum when the qubits are in the even subspace and will become slightly vacuum-squeezed under the action of the off-resonant two-photon drive. The axis of squeezing, schematically represented in the inset of \cref{fig:schematic}\textbf{b}, depends on the sign of the parametric pump detuning, and is therefore different for the two even states $\ket{00}$ and $\ket{11}$. This results in slow dephasing within the even parity subspace at rate $\gamma_\mathrm{e} = \kappa(\mEp/2\chi)^2$ (see Methods). This dephasing can be made small by limiting the amplitude of the two-photon drive $\mEp/2\chi \ll 1$. Crucially, this does not limit the SNR of the measurement since $|\alpha_\mathrm{o}|$ can be made large by reducing the resonator nonlinearity, $K$, as shown by \cref{eq:alpha0amplitude}. In other words, the measurement SNR and the eigenspace dephasing rate $\gamma_e$ can be optimized separately.
This is in stark contrast with schemes based on coherent drives where, for a fixed dispersive coupling $\chi$, the eigenspace dephasing increases with the SNR~\cite{Lalumiere:10a,Hutchison:09a,Tornberg:10a}. 

\begin{figure}[!t]
\centering
\includegraphics[scale = 1]{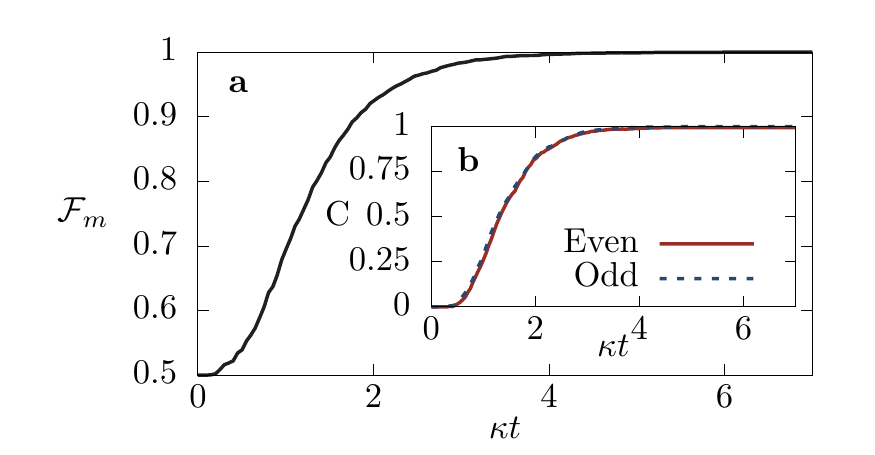}
\caption{\textbf{a} Measurement fidelity as a function of time. \textbf{b} Concurrence conditioned on the measurement record being even (red) or odd (blue). The parameters are $K/\kappa = 0.17$, $\chi/\kappa = 25$ and $\mEp/\kappa = 2.5$ for both panels.}
\label{fig:fidelity}
\end{figure}

To numerically evaluate the performance of this measurement scheme, we simulate the evolution of \cref{eq:ham2qb} under a stochastic master equation (see Methods)~\cite{Wiseman:10a}.
We first compute 2000 trajectories where the qubits are initialized in the odd (even) subspace. For each trajectory, we integrate the resulting homodyne current and categorize it as odd (even) if the absolute value of the signal is above (below) an optimized threshold value.
The resulting measurement fidelity $\mathcal F_m(\tau) = 1/2[P(e|e) + P(o|o)]$ is shown as a function of time in \cref{fig:fidelity}\textbf{a}. Starting at $\mathcal F_m(0) = 0.5$ corresponding to a random parity guess, the fidelity steadily increases towards 1. For the realistic parameters $K/\kappa = 0.175$, $\chi/\kappa = 25$, $\mEp/\kappa = 2.5$ and $\tau = 5/\kappa$, we find a large measurement fidelity $\mathcal F_m = 99.9\%$. In these simulations, the steady state photon number (in the odd subspace) is set to $|\alpha_\mathrm{o}|^2 \approx 14$, leading to a high SNR once the resonator reaches steady state. For these parameters the measurement time is thus limited by the bifurcation time to the steady state, which scales as $\sim 1/(\mEp - \kappa/2)$ (see Methods). This could potentially be shortened by shaping the two-photon pulse $\mEp(t)$ or with further parameter optimization. Moreover, the measurement fidelity might be improved further by using more sophisticated signal analysis methods such as machine learning techniques~\cite{Magesan:15a}. 

Starting with an unentangled superposition of the odd and even states, this parity measurement collapses the qubits to an entangled Bell state within one of the two subspaces. To study the creation of entanglement and assess the importance of eigenspace dephasing, we initialize the system in an unentangled state with both qubits in the +1 eigenstate of $\sx$ and the resonator in the vacuum state, $\ket{++}\otimes\ket{0}$. We again compute 2000 realizations of the evolution and register the qubits state conditioned on the measurement record, $\rho_c$. \Cref{fig:fidelity}\textbf{b} shows the 
concurrence of $\rho_c$ as a function of the measurement time $\tau$. From the initial unentangled state, the qubits are rapidly projected on one of the two parity subspaces, leading to a high concurrence at moderate times. At longer times, the concurrence conditioned on an odd parity measurement approaches unity and, in the even subspace, it slowly decreases due to the slow dephasing $\gamma_\mathrm{e}$ (not apparent on the scale of \cref{fig:fidelity}\textbf{b}). In order to study the properties of the measurement process only, we considered ideal qubits ($T_1,T_2 \rightarrow \infty$) and perfectly matched dispersive shifts.
In practice, these imperfections will cause the concurrence to slowly decrease and, in the case of relaxation errors during the measurement ($T_1$), will decrease the measurement fidelity $\mathcal F_m$.

After the measurement, the resonator is reset by turning off the two-photon drive and waiting for a few resonator lifetimes $1/\kappa$ or, alternatively, by adiabatically ramping down the parametric drive~\cite{Puri:17a}. Since the resonator ends up in a state close to a coherent state, this process can also be sped up using active reset techniques~\cite{Bultink:16a,McClure:16a,Boutin:17a} 


\subsection{Four-qubit Parity Measurement}
We now turn to a generalization of the above approach to four qubits. This is motivated by the many QEC codes that require frequent parity measurements of more than two qubits. This is the case, for example, of the surface code which relies on four-qubit parity measurements~\cite{Kitaev:03a}. Because of the larger Hilbert space, it is now challenging to extract the measurement fidelity and study the entangle creation from numerical simulations. As a result, in this section, we focus on the underlying concepts and on analytical results.

\begin{figure*}[!t]
\centering
\includegraphics[scale = 1]{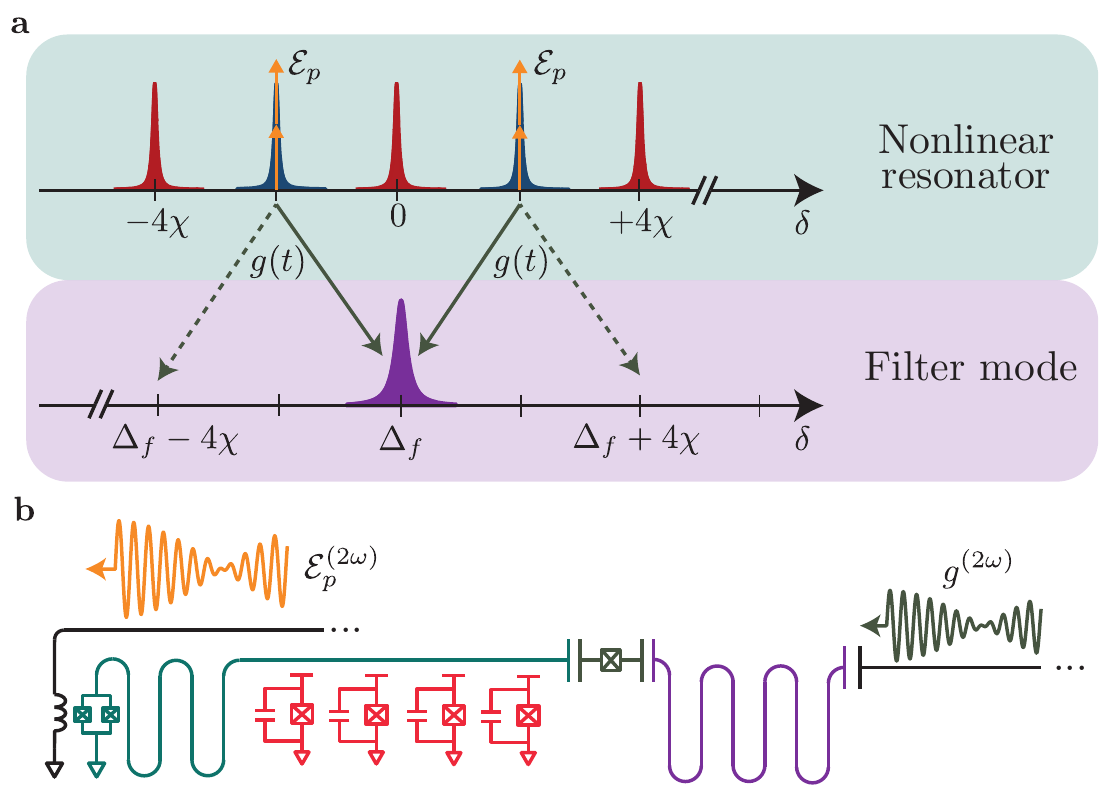}
\caption{\textbf{a} Top: Nonlinear resonator qubit-state dependent frequency. A two-tone two-photon drive $\mEp^{(2\omega)}$ is sent to the resonator at $\delta = \pm 2\chi$ (orange double arrows). Bottom: resonator photons are converted to a filter frequency (purple) via a two-tone coupling modulation $g(t)$ (dark green). \textbf{b} Possible circuit QED realization. Transmon qubits (red) are capacitively coupled to a high-Q, nonlinear resonator (light green) which is coupled via a tunable coupler (dark green) to a low-Q filter mode (purple). A two-tone microwave drive on the nonlinear resonator (orange) induces the two-photon drive while the coupling modulation is induced by the combination of a drive on the nonlinear resonator and a two-tone drive on the filter mode (dark green).}
\label{fig:schematic4qb}
\end{figure*}

Building on the results for two qubits scenario presented above, we now consider four qubits dispersively coupled to a single nonlinear resonator, where we aim to distinguish between two parity subspaces which are eightfold degenerate. In the even subspace, the dispersive shift can take three different values $\delta_\mathrm{e} = 0, \pm 4\chi$ (blue lorentzians) while, in the odd subspace, it can take two different values $\delta_\mathrm{o} = \pm 2\chi$ (red lorentzians) as schematically illustrated in \cref{fig:schematic4qb}\textbf{a}. Accordingly, a naive generalization of the two-qubit scheme presented above is to excite the resonator with a two-tone two-photon drive $\mEp^{(2\omega)}$ at frequencies $2(\wc \pm 2\chi)$, as shown by the two sets of orange double arrows in \cref{fig:schematic4qb}\textbf{a}.
As in the two-qubit case, this two-tone drive leads to a situation where the parity information is encoded in the amplitude of the resonator field: a high amplitude corresponds to the odd subspace and a null amplitude to the even subspace. When the two tones of the two-photon drive are of equal amplitude, the amplitude of the output field does not depend on the two possible dispersive shifts within the odd subspace $\delta_\mathrm{o} = \pm 2\chi$. However, the \emph{frequency} of the output field directly depends on $\delta_\mathrm{o}$, leading to fast dephasing inside the odd parity subspace at a rate $\gamma_o = \kappa |\alpha_\mathrm{o}|^2$.
A possible solution introduced for linear drive schemes~\cite{Govia:15a} and also applicable here is to use such a two-tone drive $\mEp^{(2\omega)}$ in combination with a detector that is sensitive exclusively to the amplitude of the output field, \textit{i.e.} a broadband, high-efficiency photon detector. However, the realization of this type of detector in the microwave domain remains challenging. 
Alternative proposals also offer solutions to this frequency distinguishability problem, but at the cost of higher experimental complexity~\cite{DiVincenzo:13a,Nigg:13a,Tornberg:14a,Blumoff:16a,Criger:16a}.

Here, we introduce a simpler, hardware-efficient approach to four-qubit parity measurements where the nonlinear resonator is coupled to a low-Q, ``filter'' resonator of frequency $\omega_f$ through a tunable coupling element. As we show, this effectively implements a ``frequency erasure'' channel that converts resonator photons at $\wc \pm 2\chi$ to a single frequency $\omega_f$. As a result, only the parity information remains in the output field, \textit{i.e.}~the output field contains no information about the different dispersive shifts $\delta_o$ within the odd subspace. Crucially, this allows to infer multi-qubit parity using standard homodyne detection without inducing dephasing within that subspace. 

In order to implement this frequency erasure channel, we consider a two-tone modulation $g^{(2\omega)}$ of the resonator-filter coupling at frequencies $\Delta_f \pm 2\chi$, where $\Delta_f \equiv \wc - \omega_f$.
This multi-tone coupling modulation is schematically illustrated in \cref{fig:schematic4qb}\textbf{a} (dark green arrows), where one modulation tone (full lines) brings the $\delta = \pm 2\chi$ resonator peaks (blue) in resonance with the filter mode (purple) while the other coupling modulation tone (dashed lines) is off-resonant by $\mp 4\chi$ and has only a small effect. Irrespective of the dispersive shift $\delta = \pm 2\chi$, resonator photons are then converted to a single frequency $\omega_f$. 
In a frame rotating at $\wc \pm 2\chi$ for the resonator, $\omega_f$ for the filter resonator and neglecting for now off-resonant terms, the above situation is described by the Hamiltonian (see Methods)
\begin{equation}\label{eq:ham4qbRot}
\begin{aligned}
\hH_{4qb,\mathrm{o}}^{(\pm 2\chi)} =&\,\hH_R + \frac{g}{2}\left[ \ha \hfd + \had \hf\right],
\end{aligned}
\end{equation}
where $\hf$ and $\hfd$ denote the annihilation and creation operators of the filter mode, respectively.

\Cref{eq:ham4qbRot} crucially shows that the resonators' dynamic does not depend on the state of the qubits within the odd qubit subspaces, $\delta_\mathrm{o} = \pm 2\chi$.
Consequently, similarly to the two-qubit case, four-qubit parity information can be inferred without eigenspace dephasing by monitoring the amplitude of the output field of the filter mode using homodyne detection.

Expanding further the simple analysis leading to \cref{eq:ham4qbRot} reveals that, in the odd parity subspace, the filter also emits in a qubit-state dependent sideband $\omega_f \pm 4\chi$ as illustrated by the dashed dark green lines in \cref{fig:schematic4qb}\textbf{a}. Consequently, a small portion of the ``which-frequency'' information is present in the output field, causing a slow dephasing at a rate $\gamma_o^{eff} = \kappa_{eff} |\alpha_\mathrm{o}|^2/(1+(8\chi/\kappa_f)^2)$ inside the odd subspace where $\kappa_{eff} = g^2/\kappa_f$ (see Methods). Taking a measurement time $\kappa_{eff} \tau = 5$, a steady state photon number $|\alpha_\mathrm{o}|^2 = 10$ and a ratio $\chi/\kappa_f=20$, this leads to an approximate error probability $\gamma_0^{eff} \tau = 0.2 \% $ which is below the threshold for quantum error correction with the surface code~\cite{Raussendorf:07b}.
Internal photon loss of the nonlinear resonator at a rate $\kappa_{int}$ will also induce dephasing inside the odd subspace at a rate $\gamma_o^{int} = \kappa_{int} |\alpha_\mathrm{o}|^2$, something which should ideally be minimized.

As a side note, we mention that instead of modulating the coupling, an alternative solution leading to a similar frequency-erasure channel consists in modulating the nonlinear resonator frequency $\omega_r(t)$ and adjusting the two-photon drive accordingly. Moreover, we considered above that all qubits had the same dispersive coupling $\chi$ to the nonlinear resonator. As long as the absolute value of the dispersive coupling stays homogeneous, its sign could vary amongst the qubits, $\chi \rightarrow -\chi$, with sole consequence to exchange even and odd in the discussion above. Finally, an added advantage of introducing the filter mode is that it acts naturally as a Purcell filter for the qubits~\cite{Reed:10a}.


\subsection{Circuit QED Implementation}
Realization of the above ideas is natural in different quantum systems and, as a concrete example, we now describe a possible circuit QED~\cite{Blais:04a,Wallraff:04a} implementation with transmon qubits~\cite{Koch:07a}.

\subsubsection{Two Qubits}
\Cref{fig:schematic}\textbf{c} shows the circuit for a two-qubit parity measurement, where two transmon qubits (red) are capacitively coupled to a nonlinear quarter-wavelength resonator (green). 
Taking the transmons to be far detuned from the resonator, the qubit-resonator coupling takes the dispersive character shown in \cref{eq:ham2qb}. The dispersive couplings $\chi$ are adjusted to be of equal magnitude and we assume the
transmon qubits to be detuned from each other in order to avoid qubit-qubit interaction mediated by the resonator. 
The resonator nonlinearity $K$ is induced in part by a superconducting quantum interference device (SQUID) located at the end of the resonator and in part by a qubit-induced nonlinearity.
The two-photon drive is induced by applying a flux modulation at twice the resonator frequency (orange). In short, the circuit we propose consists in two transmon qubits dispersively coupled to a Josephson Parametric amplifier (JPA) parametrically driven above threshold and is well within reach of current experimental capabilities.

\subsubsection{Four Qubits}
\Cref{fig:schematic4qb}\textbf{b} shows a possible implementation of the four-qubit parity measurement. It consists of a nonlinear, quarter-wavelength coplanar resonator (green) capacitively coupled to four transmon qubits capacitively coupled to it (red). To erase the which-frequency information, the nonlinear resonator is coupled to a linear filter resonator (purple) by a tunable coupling element (dark green).
Multiple circuits allow for the necessary coupling modulation~\cite{Yin:13a,Pierre:14a,Flurin:15a,Pfaff:17a} and here we follow Ref.~\cite{Pfaff:17a}. With this approach, the two-tone coupling modulation $g^{(2\omega)}$ is activated by driving the linear resonator (purple) with a three-tone coherent drive on the filter mode (dark green).

\section{Conclusion}

To summarize, we have introduced a scheme for qubit parity readout exploiting the bifurcation dynamics of a nonlinear oscillator. For two qubits, this leads to a high-fidelity readout that preserves the parity eigenspaces. We also presented an extension of this scheme to the parity readout of four qubits using a multi-tone parametric drive in combination with a multi-tone modulation of the coupling between a nonlinear resonator and a filter mode. 
Both schemes have a simple circuit QED implementation which could be realized with current devices.
This work paves the way for a hardware-efficient implementation of quantum error correction codes such as the surface code in circuit QED.


\section{Methods}
\subsection{Stability of Resonator Vacuum State}
When parametrically driven on resonance, the classical equations of motion for the field quadratures of the nonlinear resonator $x = \langle \ha + \had \rangle/ 2$ and $y = -i \langle \ha - \had \rangle/ 2$ are given by
\begin{equation}
\begin{aligned}
\dot x &= K (x^2 + y^2)y + \mEp y - \frac{\kappa}{2} x,\\
\dot y &= -K (x^2 + y^2)x + \mEp x - \frac{\kappa}{2} y.
\end{aligned}
\end{equation}
Computing the eigenvalues of the evolution matrix linearized around vacuum $(x,y) = (0,0)$, we obtain $\lambda_\pm = \pm \mEp - \kappa/2$. Small fluctuations around vacuum will thus make the system leave this unstable point on a timescale given by $\lambda_+^{-1} = (\mEp - \kappa/2)^{-1}$.

\subsection{Dephasing in the Two-Qubit Parity Measurement}
In the odd qubit subspace, the dispersive shifts shown in \cref{eq:ham2qb} cancel out and the qubits decouple from the resonator. Consequently, there is no dephasing in that subspace. On the other hand, in the even subspace, the two-photon parametric drive leads to a qubit-state dependant resonator field. More precisely, and as schematically illustrated in the inset of \cref{fig:schematic}\textbf{b}, when the dispersive shifts are much larger than the two-photon drive and the resonator decay rate, $4\chi \gg \mEp, \kappa$, the resonator field is in the slightly squeezed state $| r \expo{i\theta} \rangle$. The squeezing parameter is $r \approx \mEp/4\chi$ and the squeezing angle $\theta \approx 0 \text{ or }\pi/2$ is qubit-state dependent~\cite{Boutin:17b}. The overlap of these squeezed pointer states is $\langle r  | r\expo{i\pi/2} \rangle = 1/\sqrt{\cosh 2r}$. The corresponding measurement-induced dephasing in this subspace is then roughly given by $\gamma_\mathrm{e} \sim \kappa (1-\langle r  | r\expo{i\pi/2} \rangle) \sim \kappa(\mEp/4\chi)^2$ for small $r$.
A more rigorous derivation of this rate can be found in the Supplemental Material.


\subsection{Simulations}
In order to model the back-action of the homodyne measurement chain, we simulate multiple realizations of the evolution of the system under the stochastic master equation~\cite{Wiseman:10a}
\begin{equation}\label{eq:stoME}
\begin{aligned}
d\rho = -i [\hH, \rho] dt + \kappa \mathcal D[\ha] dt + \sqrt \kappa \mathcal H[\ha \expo{-i \theta_\mathrm{o}}]\rho\, dW,
\end{aligned}
\end{equation}
where $\mathcal D[\ha]\bullet = \ha \bullet \had - 1/2 \{\had \ha, \bullet\}$ is the dissipation superoperator and $\mathcal H[\hat M]\bullet = \hat M \bullet + \bullet \hat M^\dag - \text{Tr}[\hat M \bullet + \bullet \hat M^\dag]\bullet$ is the homodyne measurement back-action superoperator. Moreover, $dW$ is a Wiener increment, which has statistical properties $E[dW] = 0$, $E[dW^2] = dt$ with $E[\bullet]$ denoting the ensemble average.
The results of \cref{fig:fidelity} were obtained using \cref{eq:stoME} with the Hamiltonian \cref{eq:ham2qb}.
\Cref{eq:stoME} shows that the Hamiltonian and dissipation (first two terms) are symmetric under the transformation $\ha \rightarrow -\ha$. This symmetry is broken by the homodyne measurement backaction (last term), \textit{i.e.} by conditioning the state on the measurement record. In other words, although the average displacement of the resonator is null, conditioning the state on the measurement record makes it collapse onto $\pm \alpha_\mathrm{o}$.

The homodyne current resulting from the stochastic master equation is given by $j_h(t) = \sqrt \kappa \langle \ha \expo{-i\theta_\mathrm{o}} + \had \expo{i\theta_\mathrm{o}}\rangle + dW/dt$. For a given measurement time $\tau$, the dimensionless integrated signal is given by $s(\tau) = \sqrt \kappa \int_0^\tau dt\, j_h(t)$.

\subsection{Effective Four-qubit Hamiltonian}
As mentioned in the main text, we consider four qubits dispersively coupled to a nonlinear resonator under a two-tone two-photon drive $\mEp^{(2\omega)}(t) = \mEp \cos[2(\wc - 2\chi)t] + \mEp\cos[2(\wc + 2\chi)t]$. Coupling the nonlinear resonator to a harmonic filter through a two-tone modulation $g^{(2\omega)} = g \cos[(\Delta_f + 2\chi) t] + g \cos[(\Delta_f - 2\chi) t]$, this system is described by the Hamiltonian
\begin{equation}\label{eq:ham4qb}
\begin{aligned}
\hH_{4qb} =&\, \wc \had \ha + \chi \sum_{i=1}^4 \hat \sigma_{zi}\had \ha - \frac{K}{2} \had\had\ha\ha  + \omega_f \hfd \hf\\
& + \mEp^{(2\omega)}(t)\left[\ha\ha + \had\had\right] + g^{(2\omega)}(t) \left[ \ha \hfd + \had \hf\right].
\end{aligned}
\end{equation}
For the circuit of \cref{fig:schematic4qb}, this two-tone coupling modulation is obtained by driving the filter mode with a three-tone linear drive at frequencies $\omega_{d1},\, \omega_{d2},\, \omega_{d3}$. Setting $\omega_{d1} - \omega_{d2} = \Delta_f - 2\chi$ and $\omega_{d1} - \omega_{d3} = \Delta_f + 2\chi$ results in the desired two-tone modulation as well as AC-Stark shifts of the resonator and filter mode frequencies (see Supplemental Material). 

In order to go from \cref{eq:ham4qb} to \cref{eq:ham4qbRot} of the main text, we restrict the qubits state to the one-excitation subspace spanned by $\{\ket{0001},\ket{0010},\ket{0100},\ket{1000}\}$, leading to a dispersive shift $\delta_\mathrm{o} = -2\chi$. We then go to a frame rotating at $\wc-2\chi$ for the nonlinear resonator and at $\omega_f$ for the filter mode and, neglecting fast-rotating terms, $\hH_{4qb}$ takes the form
\begin{equation}\label{eq:ham4qbm2chi}
\begin{aligned}
\hH_{4qb,\mathrm{o}}^{(-2\chi)} =&\,\frac{\mEp}{2}(\ha\ha -\frac{K}{2} \had\had\ha\ha + \had\had)+ \frac{g}{2} \left[ \ha \hfd + \had \hf\right]\\
& + \frac{\mEp}{2}\left[\expo{i 8\chi t} \ha\ha + \expo{-i 8\chi t} \had\had \right]\\
& + \frac{g}{2}\left[\expo{i 4\chi t} \ha \hfd + \expo{-i 4\chi t}\had \hf\right].
\end{aligned}
\end{equation}
The first line corresponds to the effective Hamiltonian \cref{eq:ham4qbRot}. The second line is the off-resonant two-photon drive tone and has a small effect on the resonator. The third line leads to a small photon emission in the filter sideband $\omega_f - 4\chi$ and, consequently, to a 
dephasing rate $\kappa_{eff} |\alpha_\mathrm{o}|^2/(1+(8\chi/\kappa_f)^2)$ (see Supplemental Material). 
The effective Hamiltonian $\hH_{4qb,\mathrm{o}}^{(+2\chi)}$ in the three-excitation subspace with dispersive shift $\delta_\mathrm{o} = 2\chi$ is obtained in the same way.

\begin{acknowledgments}
This research was undertaken thanks in part to funding from NSERC, the Canada First Research Excellence Fund and the Vanier Canada Graduate Scholarships. SP acknowledges financial support by the National Science Foundation under Grant DMR-1609326 and Army Research Office under Grant W911NF1410011.
\end{acknowledgments}



%

\clearpage


\section*{Supplementary material}

Supplemental Material for
"Qubit Parity Measurement by Parametric Driving in Circuit QED"

Baptiste Royer,[1] Shruti Puri,[2] and Alexandre Blais[1,3]

[1]Institut quantique and Département de Physique, Université de Sherbrooke,
2500 boulevard de l'Université, Sherbrooke, Québec J1K 2R1, Canada
[2]Department of Applied Physics, Yale University, PO BOX 208284, New Haven, CT 06511
[3]Canadian Institute for Advanced Research, Toronto, Canada

CONTENTS

I. PARAMETRICALLY DRIVEN NONLINEAR RESONATOR

We give a summary of the different steady-states of a Kerr nonlinear resonator under two-photon driving as a function of the system parameters. The Hamiltonian of the parametrically driven nonlinear resonator is

$$\hat{H}_{R,\delta} = \delta\hat{a}^\dagger\hat{a} + \frac{\mathcal{E}_p}{2}(\hat{a}\hat{a} + \hat{a}^\dagger\hat{a}^\dagger) - \frac{K}{2}\hat{a}^\dagger\hat{a}^\dagger\hat{a}\hat{a},$$ (S1)

where $\delta = \omega_r - \omega_p/2$ is the resonator-pump detuning, $\mathcal{E}_p$ is the two-photon pump amplitude and $K$ is the resonator nonlinearity. Single photon damping at a rate $\kappa$ is also taken into account. Figure S1 illustrates the different resonator steady-states, which are separated in three regions of the parameter space [S1].

First, when the two-photon drive is below the parametric oscillation threshold, $|\mathcal{E}_p| < \sqrt{\delta^2 + (\kappa/2)^2}$, the resonator has a steady-state centered at the origin. When the resonator nonlinearity $K$ can be neglected, this steady-state is characterized by the average values [S2]

$$\langle\hat{a}^\dagger\hat{a}\rangle = \frac{\mathcal{E}_p^2}{2(\delta^2 + \kappa^2/4 - \mathcal{E}_p^2)},$$
$$\langle\hat{a}\hat{a}\rangle = \frac{-\mathcal{E}_p(\delta + i\kappa/2)}{2(\delta^2 + \kappa^2/4 - \mathcal{E}_p^2)}.$$ (S2)

Here, we are only interested in configurations far from the parametric oscillation threshold where the above average values are very small. Consequently, the effect of the nonlinearity in this regime is negligible. The effect of the nonlinearity on the resonator field in the regime close to the parametric oscillation threshold was investigated in Ref. [S3]. In the white region of Fig. S1, the steady-state described by Eq. (S2) is unique. On the other hand, when the two-photon drive is above the parametric oscillation threshold

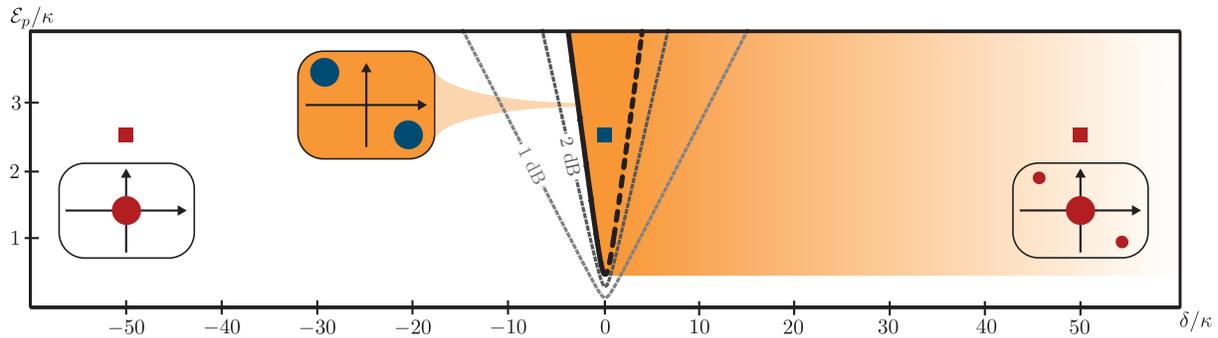

FIG. S1. Schematic representation of the steady-state of a parametrically driven nonlinear resonator in parameter space. In the white region, the resonator has a single steady-state centered at the origin. In the middle light orange region, the system has two high-amplitude steady-states. In the shaded region there are three steady-state: two high amplitude states and one state centered at the origin. The blue square indicates the configuration corresponding to parity measurement in the odd subspace, $\mathcal{E}_p/\kappa = 2.5, \delta = 0$. On the other hand, the red squares indicate the configurations in the even subspace, $\mathcal{E}_p = 2.5, \delta/\kappa = 50, \chi < 0$. The dashed gray lines correspond to 1 and 2 dB of squeezing and illustrate that for the dispersive shifts $2|\chi|/\kappa = 50$ that are considered, the level of squeezing is minimal.

$|\mathcal{E}_p| \geq \sqrt{\delta^2 + (\kappa/2)^2}$, the vacuum state becomes unstable and two new steady-states appear with average photon number

$$\langle \hat{a}^\dagger \hat{a} \rangle = \frac{\delta + \sqrt{\mathcal{E}_p^2 - (\kappa/2)^2}}{K}. \tag{S3}$$

In the resonant case $\delta = 0$, this expression corresponds to Eq. (2) of the main text. This bistable region is illustrated in orange and enclosed in the parabolic full and dashed lines in Fig. S1. Finally, when $\mathcal{E}_p > \kappa/2$ (full gray horizontal line) and $\delta > \sqrt{\mathcal{E}_p^2 - (\kappa/2)^2}$ (dashed parabolic line), both the low-amplitude steady-state described by Eq. (S2) and the two high-amplitude states described by Eq. (S3) coexist [S1]. This tristable region is shaded in light orange in Fig. S1.

As discussed in the main text, in the dispersive regime with two transmon qubits, the resonator frequency is shifted by $\pm 2\chi$ if the qubits are in the even parity subspace $\{|00\rangle, |11\rangle\}$ and remains unshifted in the odd subspace $\{|10\rangle, |01\rangle\}$. This parity-dependent frequency shift leads to an effective detuning $\delta$ with the two-photon drive that is illustrated by the blue (odd parity) and red (even parity) squares in Fig. S1. This observation is the basis for the parity measurement: even qubit parity corresponds to a low-amplitude state of the nonlinear resonator while odd qubit parity to a large amplitude state. Because the state $|00\rangle$ state places the nonlinear resonator in the tri-stability region (right red square), the resonator state can in principle tunnel from the low-amplitude state to a high-amplitude state, something that reduces the measurement fidelity and increases eigenspace dephasing in the even subspace. In practice, however, these tunneling events are highly suppressed for large dispersive shifts that we are considering, $|2\chi| \gg \mathcal{E}_p$, and can be safely neglected. Finally, we note that the overall situation described here remains unchanged in the four-qubit case and the same intuition therefore applies.

## II. DEPHASING RATES

In this section, we derive explicitly the qubit dephasing rates induced by the parity measurements for the different cases studied in the main text.

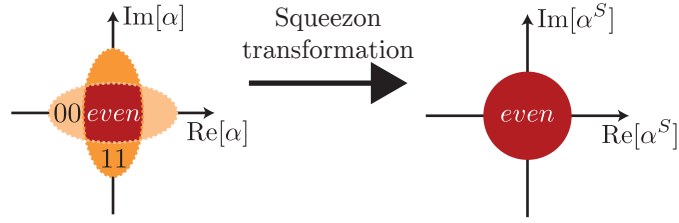

FIG. S2. Illustration of the resonator phase space when the qubits are in the even subspace. A "squeezon" transformation allows to compute the solution to the master equation in a frame where the dynamics are easier to solve.

### A. Two Qubits

The master equation for two qubits dispersively coupled to a parametrically drive nonlinear resonator is given by $(\hbar = 1)$

$$\dot{\varrho} = -i\left[\chi(\hat{\sigma}_{z1} + \hat{\sigma}_{z2})\hat{a}^\dagger\hat{a} + \frac{\mathcal{E}_p}{2}(\hat{a}\hat{a} + \hat{a}^\dagger\hat{a}^\dagger) - \frac{K}{2}\hat{a}^\dagger\hat{a}^\dagger\hat{a}\hat{a}, \varrho\right] + \kappa\mathcal{D}[\hat{a}]\varrho, \tag{S4}$$

where $\varrho$ is the combined qubits-resonator density matrix. When the qubits are in the odd subspace spanned by $\{|01\rangle, |10\rangle\}$, the above master equation reduces to

$$\dot{\varrho}_o = -i\left[\frac{\mathcal{E}_p}{2}(\hat{a}\hat{a} + \hat{a}^\dagger\hat{a}^\dagger) - \frac{K}{2}\hat{a}^\dagger\hat{a}^\dagger\hat{a}\hat{a}, \varrho_o\right] + \kappa\mathcal{D}[\hat{a}]\varrho_o, \tag{S5}$$

where $\varrho_{o(e)}$ is the qubit-resonator density matrix projected in the odd (even) qubit subspace. Since the above equation does not depend on the qubits state within the odd subspace, it is clear that the odd qubit states remain unperturbed by the measurement. In other words, there is no measurement-induced dephasing within the odd subspace.

The situation is more complicated in the even subspace spanned by $\{|00\rangle, |11\rangle\}$. In a frame rotating at $\omega_r$, the system in this subspace is described by the Hamiltonian

$$\hat{H}_{2qb,e} = 2\chi\hat{\tau}_z\hat{a}^\dagger\hat{a} + \frac{\mathcal{E}_p}{2}(\hat{a}\hat{a} + \hat{a}^\dagger\hat{a}^\dagger) - \frac{K}{2}\hat{a}^\dagger\hat{a}^\dagger\hat{a}\hat{a}, \tag{S6}$$

where we have defined $\hat{\tau}_z \equiv |11\rangle\langle11| - |00\rangle\langle00|$. Taking into account single photon loss at a rate $\kappa$, the combined qubits-resonator density matrix $\varrho_e$ evolves according to the master equation

$$\dot{\varrho}_e = -i[\hat{H}_{2qb,e}, \varrho_e] + \kappa\mathcal{D}[\hat{a}]\varrho_e. \tag{S7}$$

Following the general approach of Ref. [S4, S5], to find an expression for the measurement-induced dephasing rate in the even subspace we derive a reduced master equation for the qubit by first moving to a frame where the drive is absent. For a linear cavity and drive, this is realized with a polaron transformation, a qubit-state dependent displacement of the cavity field. In the presence of a two-photon drive, we rather introduce a "squeezon" transformation corresponding to a qubit-state dependent squeezing operation on the resonator field. As schematically illustrated in Fig. S1, this transformation is chosen such as to lead to the same resonator state irrespective of the qubit state inside in the even subspace. In this new frame the dynamics is simplified and the measurement-induced dephasing rate easier to evaluate. This unitary transformation,

$$\hat{S} = |00\rangle\langle00| \otimes e^{-\frac{r}{2}\hat{a}\hat{a} + \frac{r}{2}\hat{a}^\dagger\hat{a}^\dagger} + |11\rangle\langle11| \otimes e^{\frac{r}{2}\hat{a}\hat{a} - \frac{r}{2}\hat{a}^\dagger\hat{a}^\dagger}, \tag{S8}$$

is chosen such that $\hat{a} \equiv \hat{S}^\dagger\hat{a}\hat{S} = \cosh r\,\hat{a} - \hat{\tau}_z \sinh r\hat{a}^\dagger$, with $r$ the squeezing parameter. For the two-qubit parity measurement, we assume the dispersive shift to be much bigger than the two-photon drive, $2\chi \gg \mathcal{E}_p$, which means that the degree of squeezing is very small. Consequently, the photon population induced by the parametric drive is also very small and the nonlinearity $K$ of the resonator can be safely neglected. In the squeezon frame, the master equation governing the evolution of the density matrix $\varrho_e^S \equiv \hat{S}^\dagger\varrho_e\hat{S}$ is given by

$$\dot{\varrho}_e^S = -i[\hat{H}_{2qb,e}^S, \varrho_e^S] + \kappa\mathcal{D}[\hat{a}^S]\varrho_e \tag{S9}$$

with the Hamiltonian

$$
\begin{aligned}
\hat{H}_{2qb,e}^S &\equiv \hat{S}^\dagger \hat{H}_{2qb,e} \hat{S}, \\
&\approx (2\chi \cosh 2r - \mathcal{E}_p \sinh 2r)\hat{\tau}_z \hat{a}^\dagger \hat{a} + \left(\frac{\mathcal{E}_p}{2}\cosh 2r - \chi \sinh 2r\right)(\hat{a}\hat{a} + \hat{a}^\dagger \hat{a}^\dagger) \\
&\quad + \left(2\chi \sinh^2 r - \frac{\mathcal{E}_p}{2}\sinh 2r\right)\hat{\tau}_z, \\
&= 2\tilde{\chi}\hat{\tau}_z \hat{a}^\dagger \hat{a} + B\hat{\tau}_z,
\end{aligned}
$$

(S10)

where we have defined $2\tilde{\chi} \equiv 2\chi \cosh 2r - \mathcal{E}_p \sinh 2r$, $B \equiv 2\chi \sinh^2 r - \mathcal{E}_p/2 \sinh 2r$ to simplify the notation and set $\tanh 2r = \mathcal{E}_p/2\chi$ in order to cancel the two-photon drive. Moreover, in this frame, single-photon loss leads to the dissipators

$$
\begin{aligned}
\mathcal{D}[\hat{a}^S]\varrho_e^S &= \mathcal{D}[\hat{a}\cosh r - \hat{\tau}_z \hat{a}^\dagger \sinh r]\varrho_e^S, \\
&\approx (1 + n_{th})\mathcal{D}[\hat{a}]\varrho_e^S + n_{th}\mathcal{D}[\hat{\tau}_z \hat{a}^\dagger]\varrho_e^S,
\end{aligned}
$$

(S11)

where we have defined the effective thermal photon number in the squeezon frame $n_{th} \equiv \sinh^2 r \approx (\mathcal{E}_p/4\chi)^2$ and neglected fast oscillating terms.

The reduced master equation for the qubit density matrix $\rho_e$ is obtained by moving back to the lab frame and by tracing over the resonator,

$$
\rho_e = \mathrm{Tr}_r[\hat{S}\varrho_e^S \hat{S}^\dagger].
$$

(S12)

Following [S4], we express the combined qubits-resonator density matrix using the positive P representation [S2]

$$
\varrho_e^S = \sum_{ij=00,11} \int d^2\alpha \int d^2\beta \, P_{ij}(\alpha, \beta) \frac{|\alpha\rangle\langle\beta^*|}{\langle\beta^*|\alpha\rangle} \otimes |i\rangle\langle j|.
$$

(S13)

Back in the lab frame, the reduced qubit density matrix is then given by

$$
\rho_e = \sum_{i=00,11} \int d^2\alpha \int d^2\beta \, P_{i,i}(\alpha, \beta)|i\rangle\langle i| + \int d^2\alpha \int d^2\beta \left[\lambda(\alpha, \beta)|11\rangle\langle 00| + \lambda^*(\alpha, \beta)|00\rangle\langle 11|\right],
$$

(S14)

where we have defined

$$
\lambda(\alpha, \beta) \equiv P_{11,00}(\alpha, \beta) \frac{\langle\beta^*|e^{r\hat{a}\hat{a} - r\hat{a}^\dagger \hat{a}^\dagger}|\alpha\rangle}{\langle\beta^*|\alpha\rangle}.
$$

(S15)

Differential equation for the $P_{i,j}(\alpha, \beta)$ are obtained by using the correspondence rules [S2]

$$
\begin{aligned}
\hat{a}\varrho &\leftrightarrow \alpha P, \\
\hat{a}^\dagger \varrho &\leftrightarrow (\beta - \partial_\alpha)P, \\
\varrho\hat{a}^\dagger &\leftrightarrow \beta P, \\
\varrho\hat{a} &\leftrightarrow (\alpha - \partial_\beta)P,
\end{aligned}
$$

(S16)

taking the usual assumption that the $P$ function vanishes at infinity. The equations for the qubits diagonal elements $\dot{\rho}_{00,00} = \dot{\rho}_{11,11} = 0$ are easily solved in steady-state using a thermal state *ansatz* with average photon number $n_{th}$. On the other hand, the equation for the off-diagonal element $P_{11,00}$ reads

$$
\begin{aligned}
\dot{P}_{11,00} = \Big\{ &-2i\tilde{\chi}(2\alpha\beta - \partial_\alpha\alpha - \partial_\beta\beta) - 2iB + \frac{\kappa}{2}(\partial_\alpha\alpha + \partial_\beta\beta) \\
&-\kappa n_{th}\left[2\alpha\beta - 2(\partial_\alpha\alpha + \partial_\beta\beta) + 2 + \partial_\alpha\partial_\beta\right]\Big\} P_{11,00},
\end{aligned}
$$

(S17)

where we have defined $\partial_\gamma \equiv \partial/\partial\gamma$ ($\gamma = \alpha, \beta$). To solve the above equation, we use the transformation [S6]

$$
P(\alpha, \beta) \equiv \int da \int db \, \bar{P}(a, b)e^{i(a\alpha + b\beta) + c.c.},
$$

(S18)

for which, using integration by part, we find the identities

$$
\begin{aligned}
\alpha P(\alpha, \beta) &\rightarrow i\partial_a \bar{P}(a, b), \\
\partial_\alpha P(\alpha, \beta) &\rightarrow ia\bar{P}(a, b), \\
\partial_\alpha \alpha P(\alpha, \beta) &\rightarrow -a\partial_a \bar{P}(a, b),
\end{aligned}
\tag{S19}
$$

leading to the differential equation

$$
\begin{aligned}
\dot{\bar{P}}_{11,00} = \Big\{ &-2i\tilde{\chi}(-2\partial_a\partial_b + a\partial_a + b\partial_b) - 2iB - \frac{\kappa}{2}(a\partial_a + b\partial_b) \\
&+ 2\kappa n_{th} \left[ a\partial_a\partial_b - (a\partial_a + b\partial_b) - 1 + ab/2 \right] \Big\} \bar{P}_{11,00}.
\end{aligned}
\tag{S20}
$$

Since the equation for $\bar{P}$ is quadratic, it can be solved with a Gaussian function and we take the *ansatz*

$$
\bar{P}_{11,00}(a, b) = \mathrm{e}^\mu \mathrm{e}^{ia\bar{\alpha} + ib\bar{\beta} - \frac{a^2 X + b^2 Y}{2} - abZ},
\tag{S21}
$$

where the variables $\bar{\alpha}, \bar{\beta}, X, Y, Z$ define the resonator field and $\mu$ is a prefactor that set the phase and amplitude of the qubit density matrix off-diagonal elements $\rho_e$. Replacing this ansatz in Eq. (S20) leads to

$$
\begin{aligned}
\dot{\mu} &= -4i\tilde{\chi}(Z + \bar{\alpha}\bar{\beta}) - 2iB - 2\kappa n_{th}(Z + \bar{\alpha}\bar{\beta} + 1), \\
i\dot{\bar{\alpha}} &= 2\tilde{\chi}[2(\bar{\alpha}Z + \bar{\beta}X) + \bar{\alpha}] - i\kappa\bar{\alpha}/2 - 2i\kappa n_{th}[\bar{\alpha}Z + \bar{\beta}X + \bar{\alpha}], \\
i\dot{\bar{\beta}} &= 2\tilde{\chi}[2(\bar{\beta}Z + \bar{\alpha}Y) + \bar{\beta}] - i\kappa\bar{\beta}/2 - 2i\kappa n_{th}[\bar{\beta}Z + \bar{\alpha}Y + \bar{\beta}], \\
-\dot{X}/2 &= 2i\tilde{\chi}(2XZ + X) + \kappa X/2 + 2\kappa n_{th}(XZ + X), \\
-\dot{Y}/2 &= 2i\tilde{\chi}(2YZ + Y) + \kappa Y/2 + 2\kappa n_{th}(YZ + Y), \\
-\dot{Z} &= 4i\tilde{\chi}(XY + Z^2 + Z) + \kappa Z + 2\kappa n_{th}(XY + Z^2 + 2Z + 1/2).
\end{aligned}
\tag{S22}
$$

For simplicity, we derive an equation for $\mu$ when the resonator has reached steady-state, *i.e.* we will neglect the resonator transient dynamics and set $\dot{X} = \dot{Y} = \dot{Z} = \dot{\bar{\alpha}} = \dot{\bar{\beta}} = 0$. We easily find the steady-state solutions $X = Y = \bar{\alpha} = \bar{\beta} = 0$. Discarding the unphysical solution for Z and expanding the physical solution to first order in $n_{th}$ and $\kappa/4\tilde{\chi}$, we find

$$
Z \approx -in_{th}\frac{\kappa}{4\tilde{\chi}}.
\tag{S23}
$$

Replacing this into the equation for $\mu$ and again keeping terms to first order in $n_{th}, \kappa/4\tilde{\chi}$, we find

$$
\dot{\mu} = -\kappa n_{th} - 2iB.
\tag{S24}
$$

Using the *ansatz* for $\bar{P}_{11,00}$, we find an equation of motion for the off-diagonal elements of the reduced qubit density matrix

$$
\begin{aligned}
\dot{\rho}_{11,00} &= \int d^2\alpha \int d^2\beta \frac{\langle \beta^* | \mathrm{e}^{r\hat{a}\hat{a} - r\hat{a}^\dagger \hat{a}^\dagger} | \alpha \rangle}{\langle \beta^* | \alpha \rangle} \int da \int db\, \mathrm{e}^{i(a\alpha + b\beta) + c.c.} \dot{\bar{P}}_{11,00}(a, b), \\
&= \dot{\mu} \times \int d^2\alpha \int d^2\beta \frac{\langle \beta^* | \mathrm{e}^{r\hat{a}\hat{a} - r\hat{a}^\dagger \hat{a}^\dagger} | \alpha \rangle}{\langle \beta^* | \alpha \rangle} \int da \int db\, \mathrm{e}^{i(a\alpha + b\beta) + c.c.} \bar{P}_{11,00}(a, b), \\
&= \dot{\mu} \times \rho_{11,00}.
\end{aligned}
\tag{S25}
$$

This finally allows us to write an effective master equation for the reduced qubits density matrix which takes the form

$$
\dot{\rho} = -i[B\hat{\tau}_z, \rho] + \frac{\gamma_e}{2}\mathcal{D}[\hat{\tau}_z]\rho,
\tag{S26}
$$

with

$$
\begin{aligned}
\gamma_e &\equiv \kappa n_{th} \approx \kappa \left( \frac{\mathcal{E}_p}{4\chi} \right)^2, \\
B &\approx -\frac{\mathcal{E}_p^2}{4\chi}.
\end{aligned}
\tag{S27}
$$

In the even subspace, the qubits thus accumulate a deterministic phase $\phi = Bt$ that can be corrected with single qubit Z rotations at the end of the measurement. As discussed in the main text, the qubits also suffer from dephasing at a rate $\gamma_e$. This rate matches with the one derived in the main text in a more intuitive way.

## B.  Four Qubits

### 1.  Without Filter

We now consider a situation where four qubits are dispersively coupled to a nonlinear resonator in the presence of a two-tone, two-photon parametric drive. This situation is described by the rotating frame Hamiltonian

$$\hat{H}_{4qb,wf} = \chi \sum_{i=1}^{4} \hat{\sigma}_{zi}\hat{a}^\dagger \hat{a} + \frac{\mathcal{E}_p}{2}(e^{-i4\chi t} + e^{i4\chi t})\left[\hat{a}^\dagger \hat{a}^\dagger + \hat{a}\hat{a}\right] - \frac{K}{2}\hat{a}^\dagger \hat{a}^\dagger \hat{a}\hat{a}, \tag{S28}$$

and the master equation

$$\dot{\varrho} = -i[\hat{H}_{4qb,wf}, \varrho] + \kappa \mathcal{D}[\hat{a}]\varrho, \tag{S29}$$

where $\varrho$ is the combined qubits-resonator density matrix. In contrast to the previous subsection, here we are interested in deriving a rate $\gamma_o$ for dephasing within the odd qubit subspace, where there is no squeezing and the resonator bifurcates to states that are close to coherent states. To simplify the calculations, we take the qubits state to be within the odd subspace which is a sum of the subspace $\mathcal{H}_-$ spanned by $\{|0001\rangle, |0010\rangle, |0100\rangle, |1000\rangle\}$ with associated projector $\hat{\Pi}_-$ and the subspace $\mathcal{H}_+$ spanned by $\{|0111\rangle, |1011\rangle, |1101\rangle, |1110\rangle\}$ with associated projector $\hat{\Pi}_+$. From Eq. (S28), we see that in a given subspace $\mathcal{H}_\pm$ one parametric drive will be on resonance while the other, far detuned, will have a minimal effect if $\mathcal{E}_p/8\chi \ll 1$. Following the intuition from the two-qubit case, the combined qubit-resonator system will thus evolve to a state close to

$$|\psi_o\rangle \otimes |0\rangle \rightarrow c_+|\psi_+\rangle \otimes |\pm \alpha_o e^{-i2\chi t}\rangle + c_-|\psi_-\rangle \otimes |\pm \alpha_o e^{i2\chi t}\rangle, \tag{S30}$$

where $c_\pm|\psi_\pm\rangle \equiv \hat{\Pi}_\pm|\psi_o\rangle$ and $\alpha_o$ is given by Eq. (2) of the main text. To compute the dephasing rate $\gamma_o$, we first perform a Polaron transformation that unentangles the system and brings the resonator to a vacuum state [S5]. Then, we use the simple form of the dynamics in the Polaron frame to trace out the resonator, which allows us to write an effective master equation for the qubits where the dephasing is made explicit. This procedure is therefore similar to the previous section Sect. II A, but uses a qubit-state dependent displacement transformation instead of a qubit-state dependent squeezing transformation. Moreover, similarly to the previous section, we aim to derive a qubit dephasing rate once the resonator has reached the steady-state Eq. (S30) and we do not consider the transient dynamics.

Projecting the initial Hamiltonian Eq. (S28) onto the odd qubit subspace yields

$$\hat{H}_{4qb,wf,o} = 2\chi \hat{\tau}_z \hat{a}^\dagger \hat{a} + \mathcal{E}_p \cos(4\chi t)\left[\hat{a}^\dagger \hat{a}^\dagger + \hat{a}\hat{a}\right] - \frac{K}{2}\hat{a}^\dagger \hat{a}^\dagger \hat{a}\hat{a}, \tag{S31}$$

where $\hat{\tau}_z \equiv \hat{\Pi}_+ - \hat{\Pi}_-$. We perform a Polaron transformation $\hat{P} = \hat{\Pi}_+\hat{D}(\alpha_+) + \hat{\Pi}_-\hat{D}(\alpha_-)$, where $\hat{D}(\alpha) = \exp[\alpha\hat{a}^\dagger - \alpha^*\hat{a}]$ is the displacement operator for the resonator, $\hat{D}^\dagger(\alpha)\hat{a}\hat{D}(\alpha) = \hat{a} + \alpha$. In order to unentangle the qubits and the resonator, we choose $\alpha_\pm = \alpha_o e^{\mp i2\chi t}$, leading to the master equation

$$\dot{\varrho}_o^P = -i[\hat{H}_o^P, \varrho_o^P] + \kappa \mathcal{D}[\hat{a}]\varrho_o^P + \kappa|\delta_\alpha|^2\mathcal{D}[\hat{\tau}_z]\varrho_o^P + \kappa\delta_\alpha^*\hat{a}[\varrho_o^P, \hat{\tau}_z] + \kappa\delta_\alpha[\hat{\tau}_z, \varrho_o^P]\hat{a}^\dagger, \tag{S32}$$

with

$$\begin{aligned}
\hat{H}_o^P = {} & 2\chi\hat{a}^\dagger \hat{a}\hat{\tau}_z - \frac{K}{2}\hat{a}^\dagger \hat{a}^\dagger \hat{a}\hat{a} + 2\chi|\alpha_o|^2\hat{\tau}_z - 2K|\alpha_o|^2\hat{a}^\dagger \hat{a} \\
& + \hat{\Pi}_+ \left\{ \frac{\mathcal{E}_p}{2}\alpha_o e^{-6i\chi t}\hat{a} + \left(\frac{i\kappa}{4}\frac{\alpha_o^*}{\alpha_o}e^{4i\chi t} + \frac{\mathcal{E}_p}{2}e^{-4i\chi t}\right)\hat{a}^2 - K\alpha_o^* e^{2i\chi t}\hat{a}^\dagger \hat{a}^2 \right\} + h.c. \\
& + \hat{\Pi}_- \left\{ \frac{\mathcal{E}_p}{2}\alpha_o e^{6i\chi t}\hat{a} + \left(\frac{i\kappa}{4}\frac{\alpha_o^*}{\alpha_o}e^{-4i\chi t} + \frac{\mathcal{E}_p}{2}e^{4i\chi t}\right)\hat{a}^2 - K\alpha_o^* e^{-2i\chi t}\hat{a}^\dagger \hat{a}^2 \right\} + h.c.,
\end{aligned} \tag{S33}$$

where we have defined the displacement difference $\delta_\alpha \equiv (\alpha_+ - \alpha_-)/2 = -i\alpha_o \sin(2\chi t)$, $\varrho_o^P = \hat{P}^\dagger \varrho_o \hat{P}$ is the combined resonator-qubits density matrix in the Polaron frame and $\hat{H}_o^P = \hat{P}^\dagger \hat{H}_{4qb,wf,o}\hat{P} - i\hat{P}^\dagger \dot{\hat{P}}$ is the Hamiltonian in the Polaron frame. In the master equation Eq. (S32), we see a dephasing term proportional

to $\kappa$ (third term) appearing because of the phase difference between the displacements $\alpha_\pm$, $\delta_\alpha \neq 0$. In other words, photons leaving the resonator carry information about the qubits state, which in turn induces dephasing. The last two terms of Eq. (S32) have a minimal effect since the resonator state remains close to vacuum in the Polaron frame [S5]. In the Hamiltonian Eq. (S33), we see that the displacement of the resonator state induces an additional frequency shift of both the qubits and resonator (last two terms of the first line). Finally, the last two lines of this Hamiltonian consist of off-resonant and small terms that have a small effect on the resonator state [S7]. We can thus approximate Eqs. (S32) and (S33) with

$$\dot{\varrho}_o^P = -i[2\chi\hat{a}^\dagger\hat{a}\hat{\tau}_z - \frac{K}{2}\hat{a}^\dagger\hat{a}^\dagger\hat{a}\hat{a} + 2\chi|\alpha_o|^2\hat{\tau}_z - 2K|\alpha_o|^2\hat{a}^\dagger\hat{a}, \varrho_o^P] + \kappa\mathcal{D}[\hat{a}]\varrho_o^P + \kappa|\delta_\alpha|^2\mathcal{D}[\hat{\tau}_z]\varrho_o^P. \quad (S34)$$

In the Polaron frame, the evolution of the density matrix $\varrho_o^P$ can be calculated easily since the resonator remains in a vacuum state. To obtain a reduced master equation for the qubits density matrix $\rho_o$ in the original frame, we trace out the resonator after performing the inverse Polaron transformation

$$\rho_o = \text{Tr}_r[\hat{P}\varrho_o^P\hat{P}^\dagger]. \quad (S35)$$

Following Ref. [S5], we express $\varrho_o^P$ in the Polaron frame in the Fock basis

$$\varrho_o^P = \sum_{i,j=-,+} \sum_{n,m=0}^\infty \varrho_{i,j,n,m}^P |\psi_i, n\rangle\langle\psi_j, m|, \quad (S36)$$

and, using Eq. (S35), we write the reduced qubit density matrix as

$$\rho_o = \sum_n \varrho_{-,-,n,n}|\psi_-\rangle\langle\psi_-| + \varrho_{+,+,n,n}|\psi_+\rangle\langle\psi_+| + \sum_{n,m} \lambda_{n,m,m,n}|\psi_+\rangle\langle\psi_-| + \lambda_{m,n,n,m}^*|\psi_-\rangle\langle\psi_+|, \quad (S37)$$

where

$$\lambda_{n,m,p,q} \equiv \varrho_{+,-,n,m}^P d_{p,q} e^{-i\text{Im}[\alpha_-\alpha_+^*]}, \quad (S38)$$

$$d_{p,q} = \langle p|\hat{D}(2\delta_\alpha)|q\rangle. \quad (S39)$$

In the absence of qubit relaxation, the diagonal elements of the above expressions are simply

$$\dot{\rho}_{i,i} = \sum_n \varrho_{i,i,n,n}^P = 0. \quad (S40)$$

On the other hand, the derivative of the off-diagonal elements $\lambda$ are given by

$$\dot{\lambda}_{n,m,p,q} = \dot{\varrho}_{n,m,+,-}^P d_{p,q} e^{-i\text{Im}[\alpha_-\alpha_+^*]} - i\partial_t(\text{Im}[\alpha_-\alpha_+])\lambda_{n,m,p,q}$$
$$+ 2\dot{\delta}_\alpha\sqrt{p}\lambda_{n,m,p-1,q} - 2\dot{\delta}_\alpha^*\sqrt{q}\lambda_{n,m,p,q-1} - \partial_t(\delta_\alpha\delta_\alpha^*)\lambda_{n,m,p,q}. \quad (S41)$$

Since the resonator remains in a vacuum state in the Polaron frame, only the $\lambda_{0,0,0,0}$ element is populated [S5], leading to

$$\dot{\rho}_{+,-} = \dot{\lambda}_{0,0,0,0} = \left[-2\kappa\sin^2(2\chi t)|\alpha_o|^2 - i4\chi|\alpha_o|^2 - i4\chi|\alpha_o|^2\cos(4\chi t) - \chi|\alpha_o|^2\sin(4\chi t)\right]\lambda_{0,0,0,0}. \quad (S42)$$

Assuming that the total measurement time is much larger than the timescale set by the dispersive shift $\tau_m \gg 2\pi/\chi$, the above equation can be replaced by its average over one period $2\pi/4\chi$, leading to the reduced master equation for the qubits

$$\dot{\rho}_o = -i[2\chi|\alpha_o|^2\hat{\tau}_z, \rho_o] + \kappa|\alpha_o|^2\mathcal{D}[\hat{\sigma}_z]\rho_o/2 \quad (S43)$$

As discussed in the main text, the qubits suffer from fast dephasing $\gamma_o = \kappa|\alpha_o|^2$, motivating the need for improvements on this simple set-up.

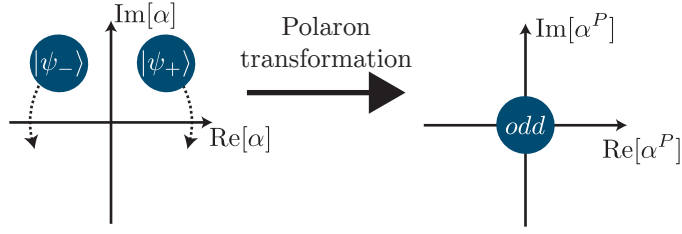

FIG. S3. Illustration of the resonator phase space when the qubits are in the odd subspace. A Polaron transformation allows to compute the solution to the master equation in a frame where the dynamics are easier to solve.

### 2. With Filter

In contrast to the previous section, we now consider that the nonlinear resonator is coupled to a harmonic mode, acting as a filter, through a two-tone modulated coupling element $g^{(2\omega)}(t) = g\cos[(\Delta_f + 2\chi)t] + g\cos[(\Delta_f - 2\chi)t]$. In a frame rotating a $\omega_r$ ($\omega_f$) for the nonlinear resonator (filter) and taking the qubits state within the odd subspace, this situation is described by the Hamiltonian

$$\hat{H}_{4qb,o} = 2\chi\hat{\tau}_z\hat{a}^\dagger\hat{a} + \mathcal{E}_p\cos(4\chi t)\left[\hat{a}^\dagger\hat{a}^\dagger + \hat{a}\hat{a}\right] - \frac{K}{2}\hat{a}^\dagger\hat{a}^\dagger\hat{a}\hat{a} + g\cos(2\chi t)(\hat{a}^\dagger\hat{f} + \hat{a}\hat{f}^\dagger),$$ (S44)

and the master equation

$$\dot{\varrho}_o = -i[\hat{H}_{4qb,o}, \varrho_o] + \kappa_f\mathcal{D}[\hat{f}]\varrho_o + \kappa_{int}\mathcal{D}[\hat{a}]\varrho_o,$$ (S45)

where $\kappa_f$ is the single photon loss rate of the filter mode and $\kappa_{int}$ is the internal single photon loss rate of the nonlinear resonator. Following the previous section, we start by applying a Polaron transformation on both the nonlinear resonator and the harmonic filter mode $\hat{P} = \hat{\Pi}_+\hat{D}_A(\alpha_+)\hat{D}_f(\beta_+) + \hat{\Pi}_-\hat{D}_A(\alpha_-)\hat{D}_f(\beta_-)$, where

$$\beta_\pm = \frac{-ig\alpha_o}{\kappa_f}\left(1 + \frac{e^{\mp i4\chi t}}{1 \mp i8\chi/\kappa_f}\right),$$ (S46)

$$\delta_\beta \equiv \frac{\beta_+ - \beta_-}{2} = \frac{g\alpha_o}{\kappa_f}\left(\frac{-\sin(4\chi t) + 8\chi/\kappa_f\cos(4\chi t)}{1 + (8\chi/\kappa_f)^2}\right).$$ (S47)

The equations for $\beta_\pm(t)$ correspond to the assymptotic solutions for $\langle\hat{f}\rangle(t)$ when taking $\langle\hat{a}\rangle = \alpha_o e^{\mp i2\chi t}$. In the Polaron frame, the master equation is given by

$$\begin{aligned}\dot{\varrho}_o^P = &-i[\hat{H}_{4qb,o}^P, \varrho_o^P] + \kappa_f\mathcal{D}[\hat{f}]\varrho_o^P + \kappa_{int}\mathcal{D}[\hat{a}]\varrho_o^P + (\kappa_f|\delta_\beta|^2 + \kappa_{int}|\delta_\alpha|^2)\mathcal{D}[\hat{\tau}_z]\varrho_o^P \\ &+ \kappa_f\delta_\beta^*\hat{f}[\varrho_o^P, \hat{\tau}_z] + \kappa_B\delta_\beta[\hat{\tau}_z, \varrho_o^P]\hat{f}^\dagger + \kappa_{int}\delta_\alpha^*\hat{a}[\varrho_o^P, \hat{\tau}_z] + \kappa_{int}\delta_\alpha[\hat{\tau}_z, \varrho_o^P]\hat{a}^\dagger,\end{aligned}$$ (S48)

with

$$\begin{aligned}\hat{H}_{4qb,o}^P = &2\chi\hat{a}^\dagger\hat{a}\hat{\tau}_z - \frac{K}{2}\hat{a}^\dagger\hat{a}^\dagger\hat{a}\hat{a} + 2\chi|\alpha_o|^2\hat{\tau}_z - 2K|\alpha_o|^2\hat{a}^\dagger\hat{a} + g\cos(2\chi t)(\hat{a}^\dagger\hat{f} + \hat{f}\hat{a}^\dagger) \\ &+ \hat{\Pi}_+\left\{\left[\frac{\mathcal{E}_p}{2}\alpha_o e^{-6i\chi t} + \frac{i\kappa_{eff}}{2}\alpha_o^*\left(e^{-2i\chi t} + e^{4i\chi t}\frac{1}{1 + i8\chi/\kappa_f}\right)\right]\hat{a} \\ &\quad + \left(\frac{i\kappa_{eff}}{4}\frac{\alpha_o^*}{\alpha_o}e^{4i\chi t} + \frac{\mathcal{E}_p}{2}e^{-4i\chi t}\right)\hat{a}^2 - K\alpha_o^*e^{2i\chi t}\hat{a}^\dagger\hat{a}^2\right\} + h.c. \\ &+ \hat{\Pi}_-\left\{\left[\frac{\mathcal{E}_p}{2}\alpha_o e^{6i\chi t} + \frac{i\kappa_{eff}}{2}\alpha_o^*\left(e^{2i\chi t} + e^{-4i\chi t}\frac{1}{1 - i8\chi/\kappa_f}\right)\right]\hat{a} \\ &\quad + \left(\frac{i\kappa_{eff}}{4}\frac{\alpha_o^*}{\alpha_o}e^{-4i\chi t} + \frac{\mathcal{E}_p}{2}e^{4i\chi t}\right)\hat{a}^2 - K\alpha_o^*e^{-2i\chi t}\hat{a}^\dagger\hat{a}^2\right\} + h.c.,\end{aligned}$$ (S49)

where we have defined $\kappa_{eff} \equiv g^2/\kappa_f + \kappa_{int}$. Similarly to the no filter case analyzed in the previous section, the two resonator modes remain close to a vacuum state in the Polaron frame, which means that the second

line in Eq. (S48) has a minimal effect and can be neglected [S5]. Moreover, the last four lines of $\hat{H}^P_{4qb,o}$ are neglected because they consist of small and rotating terms, only inducing small fluctuations. We thus approximate the master equation Eq. (S48) to

$$
\begin{aligned}
\dot{\varrho}^P_o = &- i \left[ 2\chi \hat{a}^\dagger \hat{a} \hat{\tau}_z - \frac{K}{2} \hat{a}^\dagger \hat{a}^\dagger \hat{a} \hat{a} + 2\chi |\alpha_o|^2 \hat{\tau}_z - 2K|\alpha_o|^2 \hat{a}^\dagger \hat{a} + g \cos(2\chi t)(\hat{a}^\dagger \hat{f} + \hat{f} \hat{a}^\dagger), \varrho^P_o \right] \\
&+ \kappa_f \mathcal{D}[\hat{f}] \varrho^P_o + (\kappa_f |\delta_\beta|^2 + \kappa_{int} |\delta_\alpha|^2) \mathcal{D}[\hat{\tau}_z] \varrho^P_o + \kappa_{int} \mathcal{D}[\hat{a}] \varrho^P_o.
\end{aligned}
\tag{S50}
$$

The reduced state of the effective qubit in the lab frame is given by

$$
\rho_o = \text{Tr}_r [\hat{P} \varrho^P_o \hat{P}^\dagger],
\tag{S51}
$$

where we write the state in the Polaron frame as

$$
\varrho^P_o = \sum_{n,m,l,k=0}^{\infty} \sum_{i,j=-,+} \varrho^P_{i,j,n,m,l,k} |\psi_i, n, l\rangle \langle \psi_j, m, k|,
\tag{S52}
$$

with $n, m$ $(l, k)$ referring to the nonlinear resonator (filter mode) Fock states. The reduced qubit density matrix $\rho_o$ is thus given by

$$
\rho_o = \sum_{n,l} \varrho_{-,-,n,n,l,l} |\psi_-\rangle \langle \psi_-| + \varrho_{+,+,n,n,l,l} |\psi_+\rangle \langle \psi_+| + \sum_{n,m} \lambda_{n,m,m,n,l,k,k,l} |\psi_+\rangle \langle \psi_-| + \lambda^*_{m,n,n,m,l,k,k,l} |\psi_-\rangle \langle \psi_+|,
\tag{S53}
$$

where

$$
\lambda_{n,m,p,q,k,l,r,s} = \varrho^P_{+,-,n,m,k,l} d^\alpha_{p,q} e^{-i\text{Im}[\alpha-\alpha^*_+]} d^\beta_{r,s} e^{-i\text{Im}[\gamma-\gamma^*_+]},
\tag{S54}
$$

$$
d^\alpha_{p,q} \equiv \langle p| D(2\delta_\alpha) |q\rangle,
\tag{S55}
$$

$$
d^\beta_{r,s} \equiv \langle r| D(2\delta_\beta) |s\rangle.
\tag{S56}
$$

The equations of motion of the reduced qubit diagonal elements are simply

$$
\dot{\rho}_{i,i} = \sum_{n,l} \varrho^P_{i,i,n,n,l,l} = 0.
\tag{S57}
$$

On the other hand, the equation of motion for the qubits off-diagonal elements is more complex and given by

$$
\begin{aligned}
\dot{\lambda}_{n,m,p,q,k,l,r,s} = &\ \dot{\varrho}^P_{+,-,n,m,k,l} d^\alpha_{p,q} e^{-i\text{Im}[\alpha-\alpha^*_+]} d^\beta_{r,s} e^{-i\text{Im}[\gamma-\gamma^*_+]} \\
&- i\partial_t (\text{Im}[\alpha-\alpha_+]) \lambda_{n,m,p,q,k,l,r,s} + \dot{\delta}_\alpha \sqrt{p} \lambda_{n,m,p-1,q,k,l,r,s} \\
&- \dot{\delta}^*_\alpha \sqrt{a} \lambda_{n,m,p,q-1,k,l,r,s} - 2\partial_t (\delta_\alpha \delta^*_\alpha) \lambda_{n,m,p,q,k,l,r,s} \\
&- i\partial_t (\text{Im}[\beta-\beta_+]) \lambda_{n,m,p,q,k,l,r,s} + \dot{\delta}_\beta \sqrt{r} \lambda_{n,m,p,q,k,l,r-1,s} \\
&- \dot{\delta}^*_\beta \sqrt{s} \lambda_{n,m,p,q,k,l,r,s-1} - 2\partial_t (\delta_\beta \delta^*_\beta) \lambda_{n,m,p,q,k,l,r,s}.
\end{aligned}
\tag{S58}
$$

Similarly to the previous section, we use the fact that in the Polaron frame, both nonlinear resonator and filter mode remain in a vacuum state. Consequently, only the $\lambda_{0,0,0,0,0,0,0,0,0}$ element is populated [S5]. Averaging the resulting equation over one period $2\pi/4\chi$, we get the equation

$$
\dot{\rho}_{+,-} = \dot{\lambda}_{0,0,0,0,0,0,0,0,0} = \left[ -\gamma^{eff}_o - i4\chi |\alpha_o|^2 \right] \lambda_{0,0,0,0,0,0,0,0,0},
\tag{S59}
$$

where we have defined

$$
\gamma^{eff}_o \equiv \kappa_{eff} |\alpha_o|^2 \frac{1}{1+(8\chi/\kappa_f)^2} + \kappa_{int} |\alpha_o|^2.
\tag{S60}
$$

Using these results, we finally obtain the reduced qubit master equation

$$
\dot{\rho}_o = -i[2\chi |\alpha_o|^2 \hat{\tau}_z, \rho_o] + \gamma^{eff}_o \mathcal{D}[\hat{\tau}_z] \rho_o/2.
\tag{S61}
$$

This expression is the same as found in Eq. (S43) without the filter mode, but now with a renormalized decay rate $\gamma^{eff}_o$. Here, dephasing can be kept under control by designing the system such that $8\chi/\kappa_f \gg 1$ and by reducing the internal photon loss rate of the nonlinear resonator, $\kappa_{int}$, as much as possible. Moreover, the deterministic phase proportional to $4\chi |\alpha_o|^2$ accumulated inside this subspace can be corrected by setting the measurement time so that the phase is an integer multiple of $2\pi$ or, alternatively, with a series of single-qubit phase gates.

## III. CIRCUIT QED IMPLEMENTATION

In this section, we show how the circuits illustrated in Figs. S4 and S5 implement respectively the Hamiltonians Eqs. (5) and (9) of the main text.

### A. Two qubits

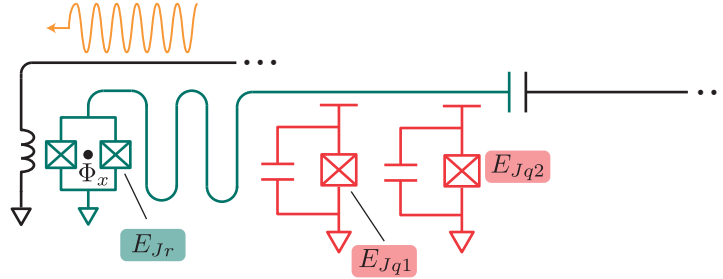

FIG. S4. Possible circuit QED implementation of the two-qubits parity measurement. Two transmon qubit (red) are dispersively coupled to a nonlinear resonator (green). The nonlinear resonator is ended by a SQUID loop threaded by a modulated flux $\Phi_x(t)$. For simplicity, we consider two identical Josephson junctions of energy $E_{Jr}$ for the resonator SQUID loop. We denote the Josephson energy of the two transmon qubits by $E_{Jq1}, E_{Jq2}$ respectively.

Using standard circuit quantization techniques [S8] and expanding the qubits Josephson junctions potential to fourth order, one finds that the circuit illustrated in Fig. S4 is well described by the Hamiltonian ($\hbar = 1$)

$$\hat{H} = \omega_0 \hat{a}^\dagger \hat{a} - E_{Jr} \cos\left[\frac{\Phi_x(t)}{2\Phi_0}\right] \cos(\hat{\varphi}_r) - E_{Jr} J_0(\delta f) \cos F \frac{\hat{\varphi}_r^2}{2} + \sum_{i=1}^{2} \left[\omega_{qi} \hat{b}_i^\dagger \hat{b}_i - \frac{E_{Jqi}}{24} \hat{\varphi}_{qi}^4 + g_i(\hat{a}\hat{b}_i^\dagger + \hat{a}^\dagger \hat{b}_i)\right].$$
(S62)

where $\omega_0, \omega_{qi}$ are the bare frequencies of the resonator and transmon qubits respectively, $\Phi_0$ is the quantum of flux, $\hat{\varphi}_r = \Phi_{zpf,r}(\hat{a} + \hat{a}^\dagger)/\Phi_0$ is the dimensionless phase difference across the resonator junctions, $J_n(x)$ is the $n^{th}$ Bessel function of the first kind and $g_i$ is the coupling strength between the resonator and the $i^{th}$ qubit. We take a flux modulation of the form $\Phi_x(t)/2\Phi_0 \equiv F + \delta f \cos(\omega_p t)$ where $F, \delta f$ are the dimensionless amplitude of the static and modulated part of the flux and $\omega_p$ is the modulation frequency. We use the Jacobi-Anger expansion to express the second term of Eq. (S62) as

$$E_{Jr} \cos(F + \delta f \cos \omega_p t) = \sum_n E_{Jr}^{(n)} \cos(n\omega_p t),$$
(S63)

where we have defined

$$
\begin{aligned}
E_{Jr}^{(0)} &\equiv E_{Jr} J_0(\delta f) \cos(F), \\
E_{Jr}^{(2n-1)} &\equiv 2E_{Jr}(-1)^n J_{2n-1}(\delta f) \sin(F), \\
E_{Jr}^{(2n)} &\equiv 2E_{Jr}(-1)^n J_{2n}(\delta f) \cos(F).
\end{aligned}
$$
(S64)

Diagonalizing the static quadratic part of the Hamiltonian, we get

$$\hat{\tilde{H}} \approx \tilde{\omega}_0 \hat{a}^\dagger \hat{a} - \sum_n E_{Jr}^{(n)} \cos(n\omega_p t) \cos(\hat{\tilde{\varphi}}_r) - E_{Jr}^{(0)} \frac{\hat{\tilde{\varphi}}_r^2}{2} + \sum_i \tilde{\omega}_{qi} \hat{b}_i^\dagger \hat{b}_i - \frac{E_{Jqi}}{24} \hat{\tilde{\varphi}}_{qi}^4,$$
(S65)

where $\hat{\tilde{\varphi}}_r = \phi_{r,r}(\hat{a} + \hat{a}^\dagger) + \sum_i \phi_{r,qi}(\hat{b}_i + \hat{b}_i^\dagger), \hat{\tilde{\varphi}}_{qi} = \phi_{qi,r}(\hat{a} + \hat{a}^\dagger) + \sum_j \phi_{qi,qj}(\hat{b}_j + \hat{b}_j^\dagger)$. $\phi_{n,m}$ denotes the dimensionless zero point fluctuations of the flux of mode $m$ across the junction of $n$ and $\tilde{\omega}_0, \tilde{\omega}_{qi}$ denote the renormalized frequencies of the resonator and transmon qubits respectively. Here, we take a small flux

modulation amplitude $\delta f \ll 1$ and use the fact that $J_n(\delta f) \approx (\delta f/2)^n/n!$ to keep only the first two terms in the sum of harmonics Eq. (S63). Expanding the resonator junction potential to fourth order, we get

$$\hat{H} \approx \tilde{\omega}_r \hat{a}^\dagger \hat{a} - E_{Jr}^{(0)} \frac{\hat{\varphi}_r^4}{24} + E_{Jr}^{(1)} \cos(\omega_p t) \frac{\hat{\varphi}_r^2}{2} + \sum_i \tilde{\omega}_{qi} \hat{b}_i^\dagger \hat{b}_i - \frac{E_{Jqi}}{24} \hat{\varphi}_{qi}^4 \tag{S66}$$

Expanding $\hat{\varphi}_r, \hat{\varphi}_{qi}$ and neglecting fast-rotating terms taking into account that we will take the flux modulation frequency at twice the resonator frequency $\omega_p = 2\omega_r$, we get

$$\hat{H} \approx (\tilde{\omega}_0 - K + \sum_i \chi_i) \hat{a}^\dagger \hat{a} - \frac{K}{2} \hat{a}^\dagger \hat{a} \hat{a} \hat{a} + \mathcal{E}_p \cos(\omega_p t)(\hat{a}\hat{a} + \hat{a}^\dagger \hat{a}^\dagger)$$

$$+ \sum_i (\tilde{\omega}_{qi} + \alpha_{qi} + \chi_i + \chi_{q1,q2}) \hat{b}_i^\dagger \hat{b}_i + \frac{\alpha_{qi}}{2} \hat{b}_i^\dagger \hat{b}_i^\dagger \hat{b}_i \hat{b}_i + 2\chi_i \hat{b}_i^\dagger \hat{b}_i \hat{a}^\dagger \hat{a} + 2\chi_{q1,q2} \hat{b}_1^\dagger \hat{b}_1 \hat{b}_2^\dagger \hat{b}_2, \tag{S67}$$

where $K \equiv E_{Jr}^{(0)} \phi_{r,r}^4/2 + \sum_i E_{Jqi} \phi_{qi,r}^4/2$ is the resonator nonlinearity, $\alpha_{qi} \equiv -E_{Jr}^{(0)} \phi_{r,qi}^4/2 - \sum_j E_{Jqi} \phi_{qj,qi}^4/2$, $\chi_i \equiv -E_{Jr}^{(0)} \phi_{r,r}^2 \phi_{r,qi}^2/2 - E_{Jqi} \phi_{qi,r}^2 \phi_{qi,qi}^2/2$ are the nonlinearity and dispersive shift of the $i^{th}$ transmon qubit respectively and $\mathcal{E}_p \equiv -E_{Jr}^{(1)} \phi_{r,r}^2/2$ is the two-photon parametric pump amplitude. Here, the qubit-qubit cross-Kerr terms $\chi_{q1,q2} = -E_{Jq1} \phi_{q1,q2}^2 \phi_{q1,q1}^2/2 - E_{Jq2} \phi_{q2,q2}^2 \phi_{q2,q1}^2/2$ are small and can be safely neglected. Projecting the transmons onto the qubit subspace $\{|0\rangle, |1\rangle\}$, we get

$$\hat{H} = \omega_r \hat{a}^\dagger \hat{a} - \frac{K}{2} \hat{a}^\dagger \hat{a} \hat{a} \hat{a} + \mathcal{E}_p \cos(\omega_p t)(\hat{a}\hat{a} + \hat{a}^\dagger \hat{a}^\dagger) + \sum_i \frac{\tilde{\omega}'_{qi}}{2} \hat{\sigma}_{zi} + \chi_i \hat{a}^\dagger \hat{a} \hat{\sigma}_{zi}, \tag{S68}$$

where we have defined $\omega_r \equiv \tilde{\omega}_0 + K + \sum_i \chi_i$, $\tilde{\omega}'_{qi} \equiv \tilde{\omega}_{qi} + \alpha_{qi} + \chi_i + \chi_{q1,q2}$. We take identical dispersive shifts $\chi_1 = \chi_2 \equiv \chi$ and, as mentioned previously, choose a flux modulation frequency $\omega_p = 2\omega_r$. After going to a frame rotating at $\omega_r$ and neglecting fast-rotating terms, we get the Hamiltonian Eq. (5) of the main text

$$\hat{H}_{2qb} = \chi(\hat{\sigma}_{z1} + \hat{\sigma}_{z2}) \hat{a}^\dagger \hat{a} + \frac{\mathcal{E}_p}{2}(\hat{a}\hat{a} + \hat{a}^\dagger \hat{a}^\dagger) - \frac{K}{2} \hat{a}^\dagger \hat{a} \hat{a} \hat{a}. \tag{S69}$$

## B. Four Qubits

Following a procedure similar to the previous section, the Hamiltonian describing Fig. S5**b** is

$$\hat{H} = \tilde{\omega}_a \hat{a}^\dagger \hat{a} - E_{Jr} \cos\left[\frac{\Phi_x(t)}{2\Phi_0}\right] \cos(\hat{\varphi}_r) - E_{Jr}^{(0)} \frac{\hat{\varphi}_r^2}{2}$$

$$+ \sum_i \tilde{\omega}_{qi} \hat{b}_i^\dagger \hat{b}_i - \frac{E_{Jqi}}{24} \hat{\varphi}_{qi}^4 + \tilde{\omega}_c \hat{c}^\dagger \hat{c} - \frac{E_{Jc}}{24} \hat{\varphi}_c^4 + \tilde{\omega}_f \hat{f}^\dagger \hat{f} + \epsilon^{(3\omega)}(t)(\hat{f} + \hat{f}^\dagger), \tag{S70}$$

where the static, quadratic, part of the Hamiltonian has already been diagonalized. Here, the transmon coupler mode (filter mode) annihilation and creation operators are denoted $\hat{c}, \hat{c}^\dagger$ ($\hat{f}, \hat{f}^\dagger$) and the phase across the junction $j$ is denoted by $\hat{\varphi}_j = \sum_{\alpha=a,c,f} \phi_{j,\alpha}(\hat{a} + \hat{a}^\dagger) + \sum_i \phi_{j,qi}(\hat{b}_i + \hat{b}_i^\dagger)$.

### 1. Flux modulation

We take a two-tone flux modulation $\Phi_x(t)/2\Phi_0 = F + \delta f_1 \cos(\omega_{p1} t) + \delta f_2 \cos(\omega_{p2} t)$. Moreover, we take small, equal modulation amplitudes for both flux modulation tones $\delta f_1 = \delta f_2 \equiv \delta f \ll 1$, leading to the first order expansion in $\delta f$

$$\cos[F + \delta f \cos(\omega_{p1} t) + \delta f \cos(\omega_{p2} t)] \approx \cos(F) J_0(\delta f)^2 - 2\sin(F) J_0(\delta f) J_1(\delta f)[\cos(\omega_{p1} t) + \cos(\omega_{p2} t)]. \tag{S71}$$

To simplify the notation, we define $E_{Jr}^{(0)} \equiv E_{Jr} J_0(\delta f)^2 \cos(F)$ and $E_{Jr}^{(1)} \equiv -2E_{Jr} J_0(\delta f) J_1(\delta f) \sin(F)$. The two-tone two-photon drive is thus given by $\mathcal{E}_p^{(2\omega)}(t) = \mathcal{E}_p \cos(\omega_{p1} t) + \mathcal{E}_p \cos(\omega_{p2} t)$, where $\mathcal{E}_p = -E_{Jr}^{(1)} \phi_{r,r}^2/2$. As schematically illustrated by the two sets of orange double arrows in Fig. S5**b**, we choose the flux modulation frequencies $\omega_{p1} = 2(\omega_r - 2\chi)$ and $\omega_{p2} = 2(\omega_r + 2\chi)$.

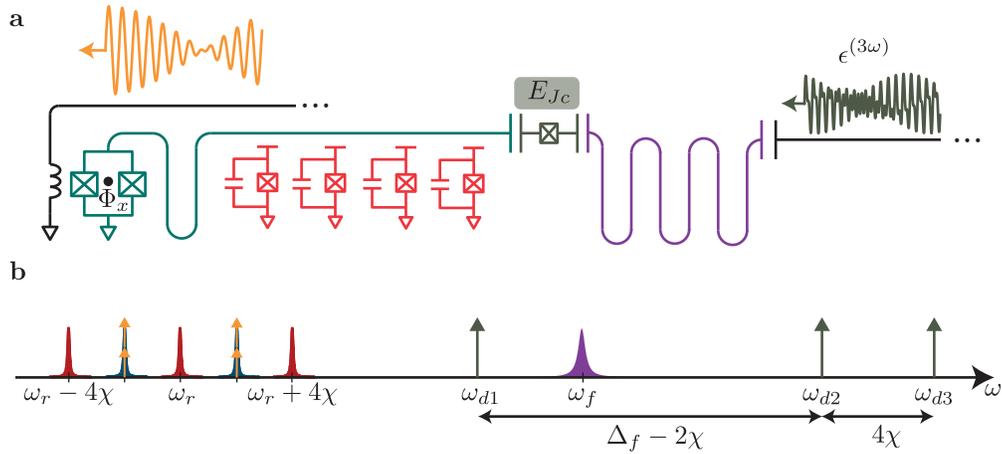

FIG. S5. **a** Possible circuit QED implementation for the 4 qubits parity measurement. Four transmon qubit (red) are dispersively coupled to a nonlinear resonator (green). The nonlinear resonator is ended by a SQUID loop threaded by a modulated flux $\Phi_x(t)$. The nonlinear resonator is capacitively coupled to a transmon mode (dark green) which in turn is capacitively coupled to a filter mode (purple). For simplicity, we consider two identical Josephson junctions of energy $E_{Jr}$ for the resonator SQUID loop. We denote the Josephson energy of the transmon qubits by $E_{Jqi}$ and the Josephson energy of the coupled transmon mode (dark green) is denoted $E_{Jc}$. Finally, a multi-tone, linear microwave drive (dark green) is applied at the input of the filter mode. **b** Illustration of the different frequencies involved in the four-qubit parity measurement. The different possible resonator frequencies when the qubits are in the even (odd) subspace are indicated by the red (blue) Lorentzians and the filter mode frequency is indicated by the purple Lorentzian. The two sets of orange double arrows indicate the parametric two-photon drive on the nonlinear resonator and the three single arrows in dark green indicate the multi-tone coherent drive on the filter mode. The drive frequencies $\omega_{di}$ are chosen such that $\omega_{d2} - \omega_{d1} = \omega_f - \omega_r - 2\chi$ and $\omega_{d3} - \omega_{d1} = \omega_f - \omega_r + 2\chi$.

### 2. Coupling Modulation

In order to induce the desired two-tone coupling modulation between the filter mode and the nonlinear resonator, we take a three-tone drive on the filter mode $\epsilon^{(3\omega)}(t) = \sum_{i=1}^{3} \epsilon_i \cos(\omega_{di} t)$. In Pfaff *et. al* [S9], a coupling between two modes was induced by driving each mode with a single-tone drive. Here, in contrast, we avoid the use of a direct drive on the nonlinear resonator as this would induce dephasing of the qubits state. Consequently, we consider a multi-tone drive that displaces the filter mode only. To understand the effect of this drive on the system, we perform the displacement transformation $\hat{f} \rightarrow \hat{f} + \sum_{i=1}^{3} \xi_i e^{-i\omega_{di} t}$, where the displacements are given by

$$\xi_i = \frac{\epsilon_i}{\Delta_i + i\kappa_f/2}, \tag{S72}$$

with the detunings $\Delta_i \equiv \omega_{di} - \omega_f$. Since the filter mode is only coupled to the coupler mode, we neglect the fluctuations of the filter normal mode across the junctions of the resonator and the data qubits, $\phi_{r,f} = \phi_{qi,f} \approx 0$. Under the displacement transformation, the nonlinear term coming from the coupler transmon mode goes to

$$-\frac{E_{Jc}}{24}\hat{\varphi}_c^4 \rightarrow -\frac{E_{Jc}}{24}\left[\sum_{\alpha=a,c,f}\phi_{c,\alpha}(\hat{\alpha}+\hat{\alpha}^\dagger) + \phi_{c,f}\sum_{i=1}^{3}(\xi_i e^{-i\omega_{di}t} + \xi_i^* e^{i\omega_{di}t})\right]^4, \tag{S73}$$

where we neglected the contributions of the data qubits normal mode to the zero point fluctuations around the coupler transmon junction, $\phi_{c,qi} \approx 0$. Keeping only the non-rotating, energy-conserving terms, we

expand Eq. (S73) to

$$
\begin{aligned}
-\frac{E_{Jc}}{24} & \left[ \sum_{\alpha=a,c,f} \phi_{c,\alpha}(\hat{\alpha} + \hat{\alpha}^\dagger) \right]^4 \\
& - E_{Jc}\phi_{c,f}^2 |\xi|^2 \sum_{\alpha=a,c,f} \phi_{c,\alpha}^2 \hat{\alpha}^\dagger \hat{\alpha} \\
& - \frac{E_{Jc}\phi_{c,f}^2}{2} \sum_{i>j} \sum_{\alpha\neq\beta=a,c,f} \phi_{c,\alpha}\phi_{c,\beta}\xi_i\xi_j \cos[(\omega_{di} - \omega_{dj})t]\hat{\alpha}^\dagger\hat{\beta},
\end{aligned}
$$

(S74)

where we have defined $|\xi|^2 \equiv \sum_i |\xi_i|^2$. The first line here corresponds to the static nonlinearity induced by the junction. The second line corresponds to an AC-stark shift of the different modes. The third line is the desired coupling modulation and, to make it resonant (in the odd data qubit subspace), we choose the drive frequencies so that $\omega_{d1} - \omega_{d2} = \omega_r - \omega_f - 2\chi$, $\omega_{d1} - \omega_{d3} = \omega_r - \omega_f + 2\chi$. Note that here $\omega_{r/f}$ have to be adjusted with the induced AC-Stark shift, something that can be done by adjusting the drive frequencies $\omega_{di}$ with the drive power. We then define $g^{(2\omega)}(t) \equiv g\cos[(\omega_{d1}-\omega_{d2})t] + g\cos[(\omega_{d1}-\omega_{d3})t]$, where $g \equiv -E_{Jc}\phi_{c,f}^3 \phi_{c,a}\xi_1\xi_2/2$. In order to make the coupling strength of equal amplitude for both modulation frequencies, the displacements associated with the drives at frequencies $\omega_{d2}, \omega_{d3}$ should be of equal magnitude $\xi_2 = \xi_3$.

### 3. Final Hamiltonian

Replacing Eqs. (S71) and (S74) in Eq. (S70) and neglecting fast-rotating terms, we get

$$
\begin{aligned}
\hat{H}' \approx{}& \omega_a'\hat{a}^\dagger\hat{a} + \mathcal{E}_p^{(2\omega)}(t)\left[\hat{a}\hat{a} + \hat{a}^\dagger\hat{a}^\dagger\right] + \omega_c'\hat{c}^\dagger\hat{c} + \omega_f'\hat{f}^\dagger\hat{f} + g^{(2\omega)}(t)\left[\hat{a}\hat{f}^\dagger + \hat{a}^\dagger\hat{f}\right] + \sum_i \omega_{qi}'\hat{b}_i^\dagger\hat{b}_i \\
& + \sum_{\alpha=a,c,f,qi} \frac{\chi_{\alpha,\alpha}}{2}\hat{\alpha}^\dagger\hat{\alpha}^\dagger\hat{\alpha}\hat{\alpha} + \sum_{\beta>\alpha=a,c,f,qi} \chi_{\alpha,\beta}\hat{\alpha}^\dagger\hat{\alpha}\hat{\beta}^\dagger\hat{\beta},
\end{aligned}
$$

(S75)

where $\chi_{\alpha,\alpha} = -\sum_\beta E_{J\beta}\phi_{\beta,\alpha}^4/2$, $\chi_{\alpha,\beta} = -\sum_\gamma E_{J\gamma}\phi_{\gamma,\alpha}^2\phi_{\gamma,\beta}^2$ and $E_{Ja} \equiv E_{Jr}^{(0)}$. The nonlinear resonator nonlinearity is given by $K = -\chi_{a,a}$ while the dispersive coupling of the qubits is given by $\chi_i = \chi_{a,qi}/2$.

We now project the data qubits onto the $\{|0\rangle, |1\rangle\}$ subspace. Moreover, since the transmon coupler mode remains in its ground state and does not play a direct role in the dynamics, it can be removed from the Hamiltonian, leading to an expression close to Eq. (9) of the main text

$$
\begin{aligned}
\hat{H}_{4qb} ={}& \omega_r\hat{a}^\dagger\hat{a} - \frac{K}{2}\hat{a}^\dagger\hat{a}^\dagger\hat{a}\hat{a} + \mathcal{E}_p^{(2\omega)}(t)\left[\hat{a}\hat{a} + \hat{a}^\dagger\hat{a}^\dagger\right] + \sum_i \frac{\omega_{qi}}{2}\hat{\sigma}_{zi} + \chi\sum_{i=1}^4 \hat{\sigma}_{zi}\hat{a}^\dagger\hat{a} \\
& + \omega_f\hat{f}^\dagger\hat{f} + g^{(2\omega)}(t)\left[\hat{a}\hat{f}^\dagger + \hat{a}^\dagger\hat{f}\right] + \frac{K_f}{2}\hat{f}^\dagger\hat{f}^\dagger\hat{f}\hat{f} + \chi_{a,f}\hat{a}^\dagger\hat{a}\hat{f}^\dagger\hat{f}.
\end{aligned}
$$

(S76)

The last two terms of the above equation lead to unwanted side effects and are a product of this particular implementation. However, $\chi_{a,f}$ can be minimized by making $\phi_{c,a} < \phi_{c,f}$ and, as long as the steady-state photon number in the filter mode Eq. (S46) is low during the parity measurement, the effect of the self-Kerr term $K_f$ is minimal.



\end{document}